%% file: main.tex
\documentclass{article}
\usepackage{graphicx}
\usepackage{color}
\usepackage{amsmath,amssymb}
\usepackage{physics}
\usepackage{hyperref}
\usepackage{tikz}
\usepackage{tikz-cd}
\usepackage{tikzit}
\usepackage{bm}
\usepackage[margin=25mm]{geometry}
\usepackage{mathtools}
\usepackage{authblk}
\numberwithin{equation}{section}

\input{sspt.tikzstyles}

\newcommand{\Bock}{\mathbf{Bock}}
\newcommand{\lattice}{\mathbf{Lattice}}
\newcommand{\dg}{\delta_{g}}
\newcommand{\df}{\delta_{f}}

\title{Lattice response theory for \\subsystem symmetry-protected topological phases}
\author{Yuki Furukawa}
\author{Takuya Okuda}
\affil{Graduate School of Arts and Sciences, University of Tokyo \\3-8-1 Komaba, Meguro-ku, Tokyo 153-8902, Japan}

\date{}

\begin{document}

\maketitle

\begin{abstract}
We develop a framework to study subsystem symmetry-protected topological (SSPT) phases using lattice response theory.
We begin with a systematic construction of spacetime lattice gauge fields from general microscopic $X$-type symmetry actions on qudit systems, including global symmetries, subsystem symmetries, and more general modulated symmetries.
Given a gauge-invariant response action, we then construct a corresponding entangler and explicitly show that it implements stacking with the SSPT phase described by the response action.
We show that a broad class of SSPT phases can be described by intersection pairings and realized by cluster entanglers.
We also study SSPT phases beyond intersection pairings, including those described by subsystem analogues of type-II group cocycles involving the Bockstein homomorphism.
We construct a new 3+1d SSPT phase protected by two 2-foliated planar subsystem symmetries and a global 0-form symmetry, whose response action is a subsystem analogue of a type-III cocycle.
On its boundary, the 0-form symmetry acts by stacking with a 2+1d linear SSPT phase, signaling a mixed 't Hooft anomaly.
\end{abstract}

\tableofcontents

\section{Introduction}

Symmetry-protected topological (SPT) phases are gapped, short-range-entangled phases of matter that become trivial when the protecting symmetry is relaxed~\cite{Chen:2010zpc,Chen:2011pg}.
Two symmetric Hamiltonians are regarded as belonging to the same phase if they can be connected by a continuous path of symmetric, gapped Hamiltonians.
For onsite symmetries, symmetric product states represent the trivial SPT phase.
Despite having no intrinsic topological order, nontrivial SPT phases can exhibit nontrivial boundary phenomena protected by symmetry.
At low energies, topological responses to background gauge fields, or equivalently symmetry defects, provide probes of the universal properties of SPT phases.
These responses are described by effective actions for background gauge fields, which define \emph{response theories}~\cite{Wang:2014pma,Tiwari:2017wqf}.
The relation between bulk responses and boundary anomalies is captured by the anomaly inflow mechanism.

A subsystem symmetry is a symmetry whose operators act on spatial subsystems.
Subsystem symmetry has been studied in connection with fracton phases~\cite{Nandkishore:2018sel,Pretko:2020cko}.
The notion of SPT phases has been extended to subsystem symmetries, giving rise to subsystem symmetry-protected topological (SSPT) phases~\cite{You:2018oai}.
Unlike ordinary global symmetries, subsystem symmetries are associated with symmetry defects whose deformations are constrained by their spatial geometry.
Consequently, such systems generally fall outside the framework of conventional relativistic field theories.
In some cases, it has also been shown that discontinuous field configurations contribute to the low-energy physics, leading to UV/IR mixing~\cite{Seiberg:2020bhn,Gorantla:2021bda}.
In particular, a conventional framework of topological field theory cannot be directly applied to SSPT phases.
Various SSPT phases have been constructed in terms of both spin systems and unconventional field theories.
However, the relation between them is not fully understood.
Moreover, certain features of SSPT phases depend on the microscopic lattice structure and are not manifest in continuum descriptions.
These observations motivate the construction of \emph{lattice response theories} that retain lattice-dependent information while connecting microscopic lattice models with continuum field theories.
This raises a natural question: How can we systematically construct background spacetime lattice gauge fields from microscopic subsystem symmetry actions in the Hamiltonian formalism and use them to formulate lattice topological responses?

In this work, we develop a general lattice framework for describing topological responses to subsystem symmetries,
establishing connections among lattice response theories, microscopic lattice Hamiltonians, and continuum field theories.
We focus on bosonic SSPT phases protected by discrete subsystem symmetries, possibly together with ordinary global symmetries.
Starting from microscopic subsystem symmetry actions, we systematically construct background spacetime lattice gauge fields, including their gauge transformations and flatness conditions.
Using these gauge fields, we construct lattice response theories for various SSPT phases.
We also show how symmetric entanglers can be constructed from these response theories, allowing us to relate the resulting topological responses to microscopic SSPT models.

We first develop a systematic method for constructing SSPT response theories based on intersection pairing.
This pairing encodes the intersections of symmetry defects.
We show that many SSPT phases previously studied in field theory can be described by this mechanism.
Furthermore, these phases admit realizations in terms of cluster Hamiltonians.

Beyond intersection pairing, we first consider 3+1d SSPT phases constructed by Devakul et al.~\cite{Devakul:2019duj}, which are protected by 2-foliated planar subsystem symmetries.
Their response theories are constructed by combining intersection pairing with the Bockstein homomorphism, using the cochain-complex structure of subsystem gauge fields.
This construction is a subsystem analogue of the response theory for conventional 2+1d type-II SPT phases protected by 0-form global symmetries.
We also construct new 3+1d SSPT phases protected by two 2-foliated planar subsystem symmetries and a 0-form global symmetry.
Their response theories are described by a subsystem analogue of the type-III cocycle for conventional 2+1d SPT phases protected by three 0-form global symmetries.
Finally, we construct a response theory for another class of 3+1d SSPT phases studied by Stephen et al.~\cite{Stephen:2020zxm},
which involves both ordinary global 0-form and planar subsystem symmetries.
We show that the symmetric entangler obtained from our response theory reproduces that of the microscopic Hamiltonian model studied in Ref.~\cite{Stephen:2020zxm}.
The response actions for the latter two examples contain cubic terms in background gauge fields, which go beyond intersection pairing.

We comment on some relations to previous work.
Procedures for gauging subsystem symmetries of lattice Hamiltonians were developed in previous studies~\cite{Williamson:2016jiq,Shirley:2018vtc}.
More recently, the gauging of finite Abelian modulated symmetries in lattice Hamiltonians was also systematically studied in~\cite{Pace:2024tgk,Yoshitome:2026qzp}.
Background \emph{spacetime} lattice gauge fields for 2+1d linear subsystem symmetry were constructed in~\cite{Cao:2022lig}.
A cochain complex for linear subsystem gauge fields was introduced in~\cite{Cao:2023rrb}.
Exactly solvable lattice models of 2+1d SSPT phases protected by linear subsystem symmetries were constructed in~\cite{You:2018oai}.
Response theories for 2+1d linear SSPT phases, which will be discussed in Section~\ref{sec:2+1d_linear_SSPT},
have been studied both in the continuum~\cite{Burnell:2021reh} and on the lattice in terms of the holonomies of background gauge fields~\cite{Cao:2023rrb,Jia:2025eff}.
These lattice responses can be expressed in terms of an intersection pairing of subsystem gauge fields, which was also used in~\cite{Cao:2023doz} as a kernel for gauging the linear subsystem symmetry.
This pairing can also be used to construct a lattice topological term in the 2+1d fractonic $\phi$-theory~\cite{Furukawa:2026okj}.\footnote{The continuum bulk theta term was first introduced in~\cite{Bedogna:2026bck}.}
Ref.~\cite{Burnell:2021reh} also constructed a 4+1d continuum SSPT phase whose boundary exhibits a 't Hooft anomaly associated with the 3+1d X-cube model, which will be discussed in Section~\ref{sec:SSPT_for_X_cube}.
Ref.~\cite{Shimamura:2024kwf} constructed 3+1d continuum SSPT response theories that cancel a mixed 't Hooft anomaly of the 2+1d exotic and foliated BF theories associated with the plaquette Ising model.
We provide a lattice description of the 2-foliated SSPT response theory in Section~\ref{sec:Shimamura_2fol}.
More generally, Ref.~\cite{Okuda:2024azp} constructed lattice SPT phases exhibiting anomaly inflow for general Calderbank--Shor--Steane (CSS) codes, including examples associated with the plaquette Ising and X-cube models.
These SPT states are called foliated cluster states for the CSS codes.
They showed that the topological responses of these phases can be expressed in terms of intersection pairings of symmetry defects.
In this work, we develop an intersection-pairing construction for SSPT phases that is not restricted to the CSS framework.
Hsin and Slagle constructed 3+1d SSPT response theories for planar subsystem symmetries using foliated two-form gauge fields~\cite{Hsin:2021mjn}.
They also formulated the responses in terms of lattice cochains and constructed corresponding exactly solvable Hamiltonians using unitary entanglers derived from the response actions.
Descriptions of SSPT phases using foliated field theories have also been studied in~\cite{Hsin:2024eyg,Gorantla:2025fzz}.
In contrast, our approach is based on lattice analogues of exotic gauge fields rather than explicit foliated gauge fields.
We also note that response field theories for dipolar SPT phases have been constructed in~\cite{Lam:2023xng}.

The rest of this paper is organized as follows.
In Section~\ref{sec:preliminaries}, we review the necessary mathematical background on (co)chain complexes.
In Section~\ref{sec:lattice_gauge_fields}, we discuss how to construct lattice gauge fields from symmetry actions and symmetric entanglers from lattice response theories.
In Section~\ref{sec:intersection_pairing}, we develop a systematic construction of lattice response theories based on intersection pairing
and apply it to several examples.
In Section~\ref{sec:beyond_intersection}, we construct lattice response theories for SSPT phases beyond intersection pairing,
including new SSPT phases in Section~\ref{sec:new_SSPT}.

\section{Preliminaries}
\label{sec:preliminaries}

\subsection{Chain and cochain complexes}

Consider a sequence of abelian groups $C_i$ for $i=0,1,\cdots,m$.
A chain complex $C_\bullet$ is defined by
\begin{align}
    C_m \xrightarrow{\partial_m} C_{m-1} \xrightarrow{\partial_{m-1}} \cdots \xrightarrow{\partial_1} C_0,
\end{align}
where $\partial_i:C_i\to C_{i-1}$ are homomorphisms satisfying $\partial_{i-1}\partial_i = 0$ for $i=2,\cdots,m$.

We define the homology groups by
\begin{align}
    H_i(C_\bullet) = \frac{\ker\partial_i}{\operatorname{im}\partial_{i+1}},
\end{align}
where $\partial_0$ and $\partial_{m+1}$ are understood to be zero maps.

For an abelian group $A$, we define the group of $A$-valued $i$-cochains by
\begin{align}
    C^i(C_\bullet;A) = \operatorname{Hom}_{\mathbb{Z}}(C_i,A),
\end{align}
for $i=0,1,\cdots,m$.
The coboundary maps $\delta_i:C^i\to C^{i+1}$ are defined by
\begin{align}
    (\delta_i f)(x) = f(\partial_{i+1}x)
\end{align}
for $f\in C^i$ and $x\in C_{i+1}$.
The associated cochain complex with coefficients in $A$ is given by
\begin{align}
    C^0(C_\bullet;A) \xrightarrow{\delta_0} C^1(C_\bullet;A) \xrightarrow{\delta_1} \cdots \xrightarrow{\delta_{m-1}} C^m(C_\bullet;A).
\end{align}
The coboundary maps satisfy $\delta_{i+1}\delta_i = 0$ for $i=0,1,\cdots,m-2$.

We define the cohomology groups by
\begin{align}
    H^i(C_\bullet;A) = \frac{\ker\delta_i}{\operatorname{im}\delta_{i-1}},
\end{align}
where $\delta_{-1}$ and $\delta_m$ are understood to be zero maps.

\paragraph{Example}
We introduce a family of two-term cochain complexes $L_{(d,N)}$ for positive integers $d$ and $N$.

Consider a $d$-dimensional hypercubic lattice with periodic boundary conditions.
Let $\mathbf{\Delta}_0$ and $\mathbf{\Delta}_1$ denote the sets of sites and $d$-dimensional cubes, respectively.
We write $s(n)$ and $c(n)$ for the generators associated with a site and a cube labeled by $n$.
We define the free abelian groups $L_{(d),i}$ generated by $\mathbf{\Delta}_i$ and consider the two-term chain complex
\begin{align}
    L_{(d),1} \xrightarrow{\partial} L_{(d),0}.
\end{align}
The boundary operator is defined by
\begin{align}
    \partial c(n) = \sum_{\sigma_{1},\cdots,\sigma_{d}\in \{0,1\}} (-1)^{d-\sum_{i=1}^d \sigma_i}s\left(n+\sum_{i=1}^{d}\sigma_{i} \hat{i}\right),
\end{align}
where $\hat{i}$ denotes a unit vector in the $i$-th direction.
We define $L_{(d,N),i} = L_{(d),i}\otimes_{\mathbb{Z}}\mathbb{Z}_N$.

The corresponding cochain complex $L_{(d,N)}$ with coefficients in $\mathbb{Z}_N$ is given by
\begin{align}
    L_{(d,N)}^0 \xrightarrow{\delta} L_{(d,N)}^1,
\end{align}
where
\begin{align}
    L_{(d,N)}^i = \operatorname{Hom}_{\mathbb{Z}}(L_{(d),i},\mathbb{Z}_N) \cong \operatorname{Hom}_{\mathbb{Z}_N}(L_{(d,N),i},\mathbb{Z}_N).
\end{align}
The coboundary operator is given by
\begin{align}
    (\delta f)(c(n)) = \sum_{\sigma_{1},\cdots,\sigma_{d}\in \{0,1\}} (-1)^{d-\sum_{i=1}^d \sigma_i}f\left(s\left(n+\sum_{i=1}^{d}\sigma_{i} \hat{i}\right)\right)
    = (\Delta_{1}\Delta_{2}\cdots \Delta_{d}f)(s(n)).
\end{align}
$\Delta_{i}$ denotes the lattice derivative along the $i$-th direction.

This cochain complex is isomorphic to the corresponding chain complex with reversed degrees.
More precisely, there exist isomorphisms
\begin{align}
    \Phi_p:L_{(d,N)}^p \to L_{(d,N),1-p}
\end{align}
satisfying
\begin{align}
    \partial\Phi_0 = \Phi_1\delta.
\end{align}
We refer to this property as \emph{self-duality}.
We use these cochain complexes in later sections.

\subsection{Tensor product of cochain complexes}

Let $C^\bullet$ and $D^\bullet$ be cochain complexes of modules over a commutative ring $R$.
Their tensor product $C^\bullet\otimes_R D^\bullet$ is defined by
\begin{align}
    (C\otimes_R D)^p = \bigoplus_{q+r=p} C^q\otimes_R D^r.
\end{align}
The coboundary operator acts on homogeneous elements as
\begin{align}
    \delta(c^q\otimes d^r) = \delta c^q\otimes d^r + (-1)^q c^q\otimes\delta d^r,
\end{align}
where $c^q\in C^q$ and $d^r\in D^r$.
The coboundary operator $\delta$ satisfies $\delta^2 = 0$.
Thus, $C^\bullet\otimes_R D^\bullet$ is again a cochain complex.

The tensor product is associative up to canonical isomorphism:
\begin{align}
    (C\otimes_R D)\otimes_R E \simeq C\otimes_R(D\otimes_R E).
\end{align}
Furthermore, the tensor product of self-dual cochain complexes of finite free $R$-modules is again self-dual.

\subsection{Pairing between chains and cochains}

Let $C_\bullet$ be a chain complex of abelian groups,
and let $C^\bullet = C^\bullet(C_\bullet;A)$ be its cochain complex with coefficients in an abelian group $A$.

There is a natural evaluation pairing
\begin{align}
    \langle\cdot,\cdot\rangle:C^p\times C_p\to A,
\end{align}
defined by
\begin{align}
    \langle\omega,\gamma\rangle = \omega(\gamma).
\end{align}
The boundary and coboundary operators satisfy
\begin{align}
    \langle\delta\omega,\gamma\rangle = \langle\omega,\partial\gamma\rangle,
\end{align}
where $\omega\in C^p$ and $\gamma\in C_{p+1}$.
Consequently, the evaluation of a cocycle on a cycle depends only on their cohomology and homology classes.
The pairing therefore descends to
\begin{align}
    \langle\cdot,\cdot\rangle:H^p(C;A)\times H_p(C)\to A.
\end{align}
In geometric settings, this pairing can be related to intersection pairings through an appropriate duality.

\subsection{Bockstein homomorphism}

Let
\begin{align}
    0 \longrightarrow B \xrightarrow{i} A \xrightarrow{p} A/B \longrightarrow 0
\end{align}
be a short exact sequence of abelian groups.
For a fixed chain complex of free abelian groups, this induces a short exact sequence of cochain complexes and hence a connecting homomorphism, called the \emph{Bockstein homomorphism},
\begin{align}
    \Bock : H^n(C_\bullet;A/B)\to H^{n+1}(C_\bullet;B).
\end{align}
It is defined as follows.

Let $\alpha\in H^n(C_\bullet;A/B)$ be a cohomology class represented by a cocycle $\omega\in Z^n(C_\bullet;A/B)$.
Choose a lift $\widetilde{\omega}\in C^n(C_\bullet;A)$ satisfying
\begin{align}
    p(\widetilde{\omega}) = \omega.
\end{align}
Since $\omega$ is closed, we have
\begin{align}
    p(\delta\widetilde{\omega}) = \delta p(\widetilde{\omega}) = \delta\omega = 0.
\end{align}
Therefore, $\delta\widetilde{\omega}$ takes values in $\ker p = \operatorname{Im}i$.
Hence there exists a unique cochain $\nu\in C^{n+1}(C_\bullet;B)$ satisfying
\begin{align}
    i(\nu) = \delta\widetilde{\omega}.
\end{align}
The Bockstein homomorphism is defined by
\begin{align}
    \Bock(\alpha) = [\nu]\in H^{n+1}(C_\bullet;B).
\end{align}
One can verify that $\nu$ is closed and that its cohomology class is independent of the choices of $\omega$ and $\widetilde{\omega}$.
Thus, $\Bock$ is a well-defined homomorphism.

\section{Spacetime lattice gauge fields}
\label{sec:lattice_gauge_fields}

\subsection{Construction of lattice gauge fields from symmetry action}
\label{sec:construction_of_gauge_fields}

In this subsection, we introduce a systematic procedure for constructing a lattice background gauge field directly from the action of a symmetry in a $\mathbb{Z}_{N}$ qudit system, or equivalently in a $\mathbb{Z}_{N}$ clock model.
We focus on on-site symmetry actions represented as products of generalized Pauli $X$-type operators.
This framework encompasses not only ordinary and higher-form global symmetries, but also subsystem symmetries and more general spatially modulated symmetries, including dipole symmetries.

\paragraph{Qudits}
Let $\Lambda$ be a finite set labeling the qudits on a spatial lattice.
For each $\lambda\in\Lambda$, the local Hilbert space $\mathcal{H}_{\lambda}$ is spanned by
$\left\{\ket{s_{\lambda}}_{\lambda}\right\}_{s_{\lambda}\in\mathbb{Z}_{N}}$
with
\begin{align}
    {}_{\lambda}\braket{s_{\lambda}}{s'_{\lambda}}_{\lambda} = \delta_{s_{\lambda},s'_{\lambda}},\qquad
    \ket{s_{\lambda}+N}_{\lambda} = \ket{s_{\lambda}}_{\lambda}.
\end{align}
We introduce the generalized Pauli operators
\begin{align}
    \mathcal{X}_{\lambda}\ket{s_{\lambda}}_{\lambda} &= \ket{s_{\lambda}+1}_{\lambda}, \\
    \mathcal{Z}_{\lambda}\ket{s_{\lambda}}_{\lambda} &= \omega^{s_{\lambda}} \ket{s_{\lambda}}_{\lambda}, \qquad
    \omega = e^{2\pi i/N}.
\end{align}
They satisfy
\begin{align}
    \mathcal{X}_{\lambda}^{N} = \mathcal{Z}_{\lambda}^{N} = 1, \qquad
    \mathcal{Z}_{\lambda}\mathcal{X}_{\lambda'} = \omega^{\delta_{\lambda,\lambda'}} \mathcal{X}_{\lambda'}\mathcal{Z}_{\lambda},
\end{align}
together with
\begin{align}
    \mathcal{X}_{\lambda}^{\dagger} = \mathcal{X}_{\lambda}^{-1}, \qquad
    \mathcal{Z}_{\lambda}^{\dagger} = \mathcal{Z}_{\lambda}^{-1}.
\end{align}

\paragraph{Local constraints and physical Hilbert space}

We define the total Hilbert space by
\begin{align}
    \mathcal{H}_{\mathrm{tot}} = \bigotimes_{\lambda\in\Lambda}\mathcal{H}_{\lambda}.
\end{align}
We allow constraints to be imposed on $\mathcal{H}_{\mathrm{tot}}$ as follows.
For $\alpha\in \operatorname{Map}(\Lambda,\mathbb{Z}_{N})$, we define
\begin{align}
    U(\alpha) = \prod_{\lambda\in\Lambda} \mathcal{X}_{\lambda}^{\alpha(\lambda)},
    \label{eq:general_X_transformation}
\end{align}
where $\operatorname{Map}(\Lambda,\mathbb{Z}_{N})$ denotes the $\mathbb{Z}_{N}$-module of maps from $\Lambda$ to $\mathbb{Z}_{N}$.
The constraints take the form
\begin{align}
    U(c)\ket{\psi} = \ket{\psi}, \qquad c\in\mathcal{C},
    \label{eq:local_constraint}
\end{align}
where $\mathcal{C}\subset \operatorname{Map}(\Lambda,\mathbb{Z}_{N})$ is a subgroup generated by elements with local support,
in the sense that the support of each generator has diameter bounded by a fixed constant independent of the system size.
We refer to these conditions as the \emph{local constraints}.
The subspace satisfying the local constraints is called the physical Hilbert space and is denoted by $\mathcal{H}_{\mathrm{phys}}\subset\mathcal{H}_{\mathrm{tot}}$.
Equivalently, on $\mathcal{H}_{\mathrm{phys}}$,
\begin{align}
    U(c)=1, \qquad c\in\mathcal{C}.
\end{align}

\paragraph{Symmetry}
We next specify the symmetry of the Hamiltonian $H$. Let $\mathcal{G}$ be a subgroup of $\operatorname{Map}(\Lambda,\mathbb{Z}_{N})$ containing $\mathcal{C}$,
and assume that
\begin{align}
    [H,U(g)] = 0, \qquad g\in\mathcal{G}.
    \label{eq:symmetry_condition}
\end{align}
Thus, $\mathcal{G}$ specifies the symmetry transformations before imposing the local constraints.
We further assume that any transformation $U(\alpha)$ commuting with every local $\mathcal{G}$-symmetric operator must satisfy $\alpha\in\mathcal{G}$.
Since elements of $\mathcal{C}$ act trivially on the physical Hilbert space, the faithfully acting symmetry is more precisely $\mathcal{G}/\mathcal{C}$.
The local constraints $\mathcal{C}$ allow us to deform the symmetry operators without changing their action on physical states, since $U(\alpha+c) = U(\alpha)$ for $\alpha \in \operatorname{Map}(\Lambda,\mathbb{Z}_{N}), c \in \mathcal{C}$ on the physical Hilbert space $\mathcal{H}_{\mathrm{phys}}$.
We call the symmetry $\mathcal{G}$ \emph{maximally topological}
if every $g \in \mathcal{G}$ with local support is trivial on the physical Hilbert space, i.e., $g \in \mathcal{C}$.\footnote{Higher-form symmetries whose symmetry operators depend on the detailed shape of their support have been discussed in \cite{Seiberg:2019vrp,Qi:2020jrf,Choi:2024rjm,Gorantla:2024ocs,Liu:2026tcl}.
They are sometimes called non-topological higher-form symmetries \cite{Choi:2024rjm,Gorantla:2024ocs,Liu:2026tcl}.
Imposing appropriate local constraints can render such symmetry operators topological, and the corresponding topological character may also emerge effectively at low energies.
In this paper, we impose the relevant local constraints exactly by hand.
See also~\cite{Liu:2026tcl} for a recent discussion of when lattice higher-form symmetries become topological at low energies.}\footnote{
Our definition of maximal topologicality excludes topological $d$-form symmetries in $d$ spatial dimensions,
since their symmetry operators have local support.}

\paragraph{Local symmetric operators}

We next characterize the local operators that can appear in a symmetric Hamiltonian.
Here, an operator is called local if its support is contained in a region whose diameter is bounded by a fixed constant independent of the system size.
When we consider a family of local operators, the bound is understood to be uniform over the family.

Since all $\mathcal{X}_{\lambda}$ commute with $U(g)$, the operators $\mathcal{X}_{\lambda}$ are automatically symmetric.
A general $\mathcal{Z}$-type monomial can be written as
\begin{align}
    \mathcal{Z}^{(p)} = \prod_{\lambda\in\Lambda} \mathcal{Z}_{\lambda}^{p(\lambda)}, \qquad p\in \operatorname{Map}(\Lambda,\mathbb{Z}_{N}).
\end{align}
Using the generalized Pauli algebra, we find
\begin{align}
    U(g) \mathcal{Z}^{(p)} U(g)^{\dagger} = \omega^{-\sum_{\lambda\in\Lambda}p(\lambda)g(\lambda)} \mathcal{Z}^{(p)}.
\end{align}
Therefore, $\mathcal{Z}^{(p)}$ is symmetric if and only if
\begin{align}
    \sum_{\lambda\in\Lambda} p(\lambda)g(\lambda) = 0 \mod N \qquad \text{for all } g\in\mathcal{G}.
    \label{eq:symmetric_Z_condition}
\end{align}

We choose a (generally non-minimal) generating set of local symmetric $\mathcal{Z}$-type monomials.
Let $\hat{\Lambda}$ denote the set labeling these generators.
We assume that every local symmetric $\mathcal{Z}$-type monomial can be expressed as a product of these generators
whose supports lie within a bounded neighborhood of the support of the monomial, independently of the system size.
We write these generators as
\begin{align}
    \mathcal{Z}^{(\hat{\lambda})} = \prod_{\lambda\in\Lambda} \mathcal{Z}_{\lambda}^{M_{\hat{\lambda},\lambda}}, \qquad \hat{\lambda}\in\hat{\Lambda},
    \label{eq:Z_generator_M}
\end{align}
where $M_{\hat{\lambda},\lambda}\in\mathbb{Z}_{N}$.
The matrix $M$ encodes the response of the local symmetric operators to arbitrary $\mathcal{X}$ transformations.
Indeed, for $\alpha\in \operatorname{Map}(\Lambda,\mathbb{Z}_{N})$,
\begin{align}
    U(\alpha) \mathcal{Z}^{(\hat{\lambda})} U(\alpha)^{\dagger} = \omega^{-(M\alpha)_{\hat{\lambda}}} \mathcal{Z}^{(\hat{\lambda})},
    \label{eq:M_action}
\end{align}
where
\begin{align}
    (M\alpha)_{\hat{\lambda}} := \sum_{\lambda\in\Lambda} M_{\hat{\lambda},\lambda} \alpha(\lambda).
\end{align}
Since $\mathcal{Z}^{(\hat{\lambda})}$ is symmetric under every $g\in\mathcal{G}$, we have
\begin{align}
    Mg = 0, \qquad g\in\mathcal{G}.
    \label{eq:M_annihilates_G}
\end{align}
Conversely, the assumptions above imply that any $\alpha$ satisfying $M\alpha=0$ belongs to $\mathcal{G}$.
Thus, $\mathcal{G}=\ker M$.

By construction, any local symmetric operator can be expressed locally in terms of $\mathcal{X}_{\lambda}$ for $\lambda\in\Lambda$ and $\mathcal{Z}^{(\hat{\lambda})}$ for $\hat{\lambda}\in\hat{\Lambda}$.
Accordingly, a general local symmetric Hamiltonian can be written schematically as
\begin{align}
    H = H\left(\mathcal{X}_{\lambda},\mathcal{Z}^{(\hat{\lambda})}\right).
    \label{eq:symmetric_Hamiltonian_general}
\end{align}
However, the expression of $H$ in terms of these generators is generally not unique due to relations among them.
This ambiguity will play an important role in the construction of defect Hamiltonians.

\paragraph{Symmetry defects in the Hamiltonian picture}

Before constructing the lattice gauge field, we review how symmetry defects appear in the Hamiltonian formalism. See Refs.~\cite{Cheng:2022sgb,Seifnashri:2023dpa} for further details.

We work in Euclidean spacetime and refer to the Euclidean-time direction simply as the temporal direction.
Consider the Euclidean partition function
\begin{align}
    Z = \operatorname{Tr}_{\mathcal{H}_{\mathrm{phys}}} \left(e^{-\beta H}\right),
\end{align}
where $\beta$ is the extent of the temporal direction.
For $g\in\mathcal{G}$,
we may insert the symmetry operator $U(g)$ at an intermediate Euclidean time $\tau_{0}$:
\begin{align}
    Z[g;\tau_{0}] = \operatorname{Tr}_{\mathcal{H}_{\mathrm{phys}}} \left(e^{-(\beta-\tau_{0})H}U(g)e^{-\tau_{0}H}\right).
    \label{eq:symmetry_insertion_partition_function}
\end{align}
In the Euclidean path-integral picture, this insertion represents a symmetry defect supported on the spatial slice $\tau=\tau_{0}$.
Since $[H,U(g)]=0$, the symmetry insertion can be moved in the temporal direction without changing correlation functions, as long as no other operator insertions are crossed.

The local constraints also allow us to deform the shape of the symmetry operator $U(g)$ without changing the partition function.
In fact, for a locally supported $c\in\mathcal{C}$,
\begin{align}
    U(g+c) = U(g)U(c) = U(g)
\end{align}
on the physical Hilbert space $\mathcal{H}_{\mathrm{phys}}$.
In a geometric representation of the defect, this relation can be interpreted as a deformation of the symmetry operator within the spatial slice $\tau=\tau_{0}$.

We next consider symmetry defects extending in the temporal direction.
Let
\begin{align}
    \alpha \in \operatorname{Map}(\Lambda,\mathbb{Z}_{N})
\end{align}
be an arbitrary transformation, which need not belong to $\mathcal{G}$.
In general, $U(\alpha) H U(\alpha)^{\dagger} \neq H$.
Using Eq.~\eqref{eq:M_action}, however, its action can be written as
\begin{align}
    U(\alpha) H U(\alpha)^{\dagger} = H\left(\mathcal{X}_{\lambda}, \omega^{-(M\alpha)_{\hat{\lambda}}}\mathcal{Z}^{(\hat{\lambda})} \right).
    \label{eq:defect_Hamiltonian}
\end{align}

Suppose, for example, that $\alpha$ agrees with a symmetry transformation in a spatial region and vanishes outside the region.
Since the corresponding untruncated symmetry transformation belongs to $\mathcal{G}$ and is annihilated by $M$,
the quantity $(M\alpha)_{\hat{\lambda}}$ is supported only near the boundary of the region.
Hence only the Hamiltonian terms near this boundary are modified.
We refer to the resulting Hamiltonian as a defect Hamiltonian, and the modified terms describe symmetry defects extending in the temporal direction.

The defect Hamiltonians obtained in this way are, however, of a special form: they are unitarily equivalent to the original defect-free Hamiltonian.
To describe more general symmetry-defect configurations, we therefore promote the phase factors appearing in the defect Hamiltonian to independent variables.
We introduce
\begin{align}
    A_s \in \operatorname{Map}(\hat{\Lambda},\mathbb{Z}_{N})
\end{align}
and define
\begin{align}
    H[A_s] := H\left(\mathcal{X}_{\lambda}, \omega^{-A_s(\hat{\lambda})}\mathcal{Z}^{(\hat{\lambda})}\right).
    \label{eq:general_defect_Hamiltonian}
\end{align}
We identify $A_s$ as the spatial component of the background gauge field.

However, this expression is generally ambiguous due to relations among the generators labeled by $\hat{\Lambda}$.
Suppose that a \emph{locally} supported $r\in\operatorname{Map}(\hat{\Lambda},\mathbb{Z}_{N})$ satisfies
\begin{align}
    \prod_{\hat{\lambda}\in\hat{\Lambda}}\left(\mathcal{Z}^{(\hat{\lambda})}\right)^{r(\hat{\lambda})} = 1.
    \label{eq:local_relation_Z_generators}
\end{align}
Here, the spatial support of $r$ is defined as the union of the supports of the generators $\mathcal{Z}^{(\hat{\lambda})}$ for which $r(\hat{\lambda})\neq 0$.
We define $\mathcal{R}$ to be the subgroup of $\operatorname{Map}(\hat{\Lambda},\mathbb{Z}_{N})$
generated by all locally supported $r$ satisfying Eq.~\eqref{eq:local_relation_Z_generators}.
Using Eq.~\eqref{eq:Z_generator_M}, Eq.~\eqref{eq:local_relation_Z_generators} is equivalent to
\begin{align}
    \sum_{\hat{\lambda}\in\hat{\Lambda}} r(\hat{\lambda}) M_{\hat{\lambda},\lambda} = 0 \mod N \qquad \text{for every } \lambda\in\Lambda.
    \label{eq:local_relation_M}
\end{align}
The existence of such relations leads to an ambiguity in the definition of $H[A_s]$.
To resolve this ambiguity, we impose
\begin{align}
    \sum_{\hat{\lambda}\in\hat{\Lambda}} r(\hat{\lambda}) A_s(\hat{\lambda}) = 0 \mod N
    \label{eq:hamiltonian_spatial_flatness}
\end{align}
for every locally supported $r$ satisfying Eq.~\eqref{eq:local_relation_M}, and hence for all $r \in \mathcal{R}$.
We call $A_s$ \emph{spatially flat} if it satisfies this condition.
Equivalently, this condition ensures that any locally supported product of generators equal to the identity remains equal to the identity after coupling to the background gauge field.
In what follows, we always impose Eq.~\eqref{eq:hamiltonian_spatial_flatness} when considering the defect Hamiltonian $H[A_s]$.

The defect Hamiltonians obtained by conjugating $H$ with $U(\alpha)$ correspond to the particular configurations $A_s = M\alpha$.
More generally, Eq.~\eqref{eq:M_action} implies
\begin{align}
    U(\alpha)H[A_s]U(\alpha)^\dagger=H[A_s+M\alpha].
\end{align}
Thus, the transformation $A_s\to A_s+M\alpha$ relates unitarily equivalent defect Hamiltonians.

We now distinguish two roles of local transformations.
Inserting a local constraint operator $U(c)$, with $c\in\mathcal{C}$, has no effect, since it acts trivially on the physical Hilbert space.
By contrast, a locally supported transformation $U(\alpha)$ relates different defect Hamiltonians through local unitary conjugation.

In the spacetime path-integral picture, both operations admit a natural interpretation as local deformations of symmetry defects, but in different directions.
The former locally deforms a space-like symmetry defect within a spatial slice.
The latter changes the location or shape of a time-like symmetry defect.

These observations provide the bridge between symmetry defects in the Hamiltonian picture and the lattice background gauge field constructed below.

\paragraph{Spacetime lattice and background gauge field}

We now discretize the temporal direction and label the time slices by $\tau\in\mathbb{Z}$.
The temporal and spatial components of the background gauge field naturally live on different sets.

We define
\begin{align}
    \mathbf{\Lambda}_{t} := \bigcup_{\tau\in\mathbb{Z}} \Lambda^{(\tau+\frac{1}{2})},
\end{align}
where $\Lambda^{(\tau+\frac{1}{2})}$ is a copy of $\Lambda$.
Its elements are denoted by $(\tau+\frac{1}{2},\lambda)$, where $\lambda\in\Lambda$.
The temporal component of the gauge field is a function
\begin{align}
    A_{t} : \mathbf{\Lambda}_{t} \longrightarrow \mathbb{Z}_{N}.
\end{align}

Similarly, define
\begin{align}
    \mathbf{\hat{\Lambda}}_{s} := \bigcup_{\tau\in\mathbb{Z}} \hat{\Lambda}^{(\tau)},
\end{align}
where $\hat{\Lambda}^{(\tau)}$ is a copy of $\hat{\Lambda}$.
Its elements are denoted by $(\tau,\hat{\lambda})$, where $\hat{\lambda}\in\hat{\Lambda}$.
The spatial component is a function
\begin{align}
    A_{s} : \mathbf{\hat{\Lambda}}_{s} \longrightarrow \mathbb{Z}_{N}.
\end{align}
We denote the combined domain of the background gauge field by
\begin{align}
    \lattice := \mathbf{\Lambda}_{t} \cup \mathbf{\hat{\Lambda}}_{s}.
\end{align}

For later use, we also introduce copies of $\Lambda$ at integer Euclidean times,
\begin{align}
    \mathbf{\Lambda}_{0} := \bigcup_{\tau\in\mathbb{Z}} \Lambda^{(\tau)}.
\end{align}
The gauge-transformation parameter $\alpha(\tau,\lambda)$ will be defined on $\mathbf{\Lambda}_{0}$.

\paragraph{Temporal component}

The temporal component $A_{t}\left(\tau+\frac{1}{2}, \lambda\right)$ records the insertion of $\mathcal{X}$ operators between the time slices $\tau$ and $\tau+1$.
More explicitly, the configuration
\begin{align}
    A_{t}\left(\tau+\frac{1}{2}\right) \in \operatorname{Map}(\Lambda,\mathbb{Z}_{N})
\end{align}
corresponds to the operator
\begin{align}
    U\left(A_{t}\left(\tau+\frac{1}{2}\right)\right) = \prod_{\lambda\in\Lambda} \mathcal{X}_{\lambda}^{A_{t}(\tau+\frac{1}{2},\lambda)}.
\end{align}

The local constraints imply a redundancy in the temporal component.
Since $U(c)=1$ on the physical Hilbert space for $c\in\mathcal{C}$,
$A_{t}\left(\tau+\frac{1}{2}\right)$ has a gauge redundancy
\begin{align}
    A_{t}\left(\tau+\frac{1}{2}\right) \longrightarrow A_{t}\left(\tau+\frac{1}{2}\right) + c, \qquad c\in\mathcal{C}.
    \label{eq:temporal_C_redundancy}
\end{align}

\paragraph{Spatial component}

The spatial component records the defect Hamiltonian.
We introduce it by modifying the local symmetric operators according to
\begin{align}
    \mathcal{Z}^{(\hat{\lambda})} \longrightarrow \omega^{-A_{s}(\tau,\hat{\lambda})} \mathcal{Z}^{(\hat{\lambda})}.
\end{align}
The Hamiltonian on the time slice $\tau$ is therefore
\begin{align}
    H[A_{s}(\tau)] := H\left(\mathcal{X}_{\lambda}, \omega^{-A_{s}(\tau,\hat{\lambda})}\mathcal{Z}^{(\hat{\lambda})}\right).
    \label{eq:background_defect_H}
\end{align}
Thus, $A_{s}(\tau,\hat{\lambda})$ specifies the phase twist attached to the local symmetric operator labeled by $\hat{\lambda}$,
and hence records the configuration of time-like symmetry defects.

The relation between a local transformation and the spatial gauge field follows immediately from
Eq.~\eqref{eq:M_action}. We have
\begin{align}
    U(\alpha)H[A_{s}(\tau)]U(\alpha)^{\dagger} = H[A_{s}(\tau)+M\alpha].
    \label{eq:defect_H_gauge_equivalence}
\end{align}
Therefore, a local transformation does not in general leave a given defect Hamiltonian invariant. Rather, it maps one defect configuration to another unitarily equivalent one.
In other words, inserting $U(\alpha)$ and $U(\alpha)^\dagger$ at times $\tau+\frac{1}{2}$ and $\tau-\frac{1}{2}$, respectively,
is equivalent to shifting the spatial component as $A_s(\tau)\to A_s(\tau)+M\alpha$.
As we will see below, this operation can be regarded as a gauge transformation in the spacetime formulation.

\paragraph{Gauge transformations}

We now combine the two types of local redundancies into the gauge transformation of the spacetime background gauge field.

Let $\alpha \in \operatorname{Map}(\mathbf{\Lambda}_{0}, \mathbb{Z}_{N})$ be a spacetime-dependent transformation,
with $\alpha(\tau,\lambda)$ defined on every integer-time slice.
In addition, for each temporal link between $\tau$ and $\tau+1$,
let $c_{\tau+\frac{1}{2}} \in \mathcal{C}$ denote an independent redundancy due to the local constraints.

The background gauge field transforms as
\begin{align}
    A_{t}\left(\tau+\frac{1}{2}, \lambda \right) \longrightarrow\;&
    A_{t}\left(\tau+\frac{1}{2}, \lambda\right) + \alpha(\tau+1,\lambda) - \alpha(\tau,\lambda) - c_{\tau+\frac{1}{2}}(\lambda),
    \label{eq:gauge_transformation_temporal} \\
    A_{s}(\tau,\hat{\lambda}) \longrightarrow\;&
    A_{s}(\tau,\hat{\lambda}) + \sum_{\lambda\in\Lambda} M_{\hat{\lambda},\lambda} \alpha(\tau,\lambda).
    \label{eq:gauge_transformation_spatial}
\end{align}
The gauge transformation parametrized by $\alpha$ comes from the redundancy discussed around Eq.~\eqref{eq:defect_H_gauge_equivalence}.
On the other hand, the $c$-gauge transformation acting only on the temporal component of the gauge field is associated with the local constraints.

As a consistency check, suppose that $\alpha(\tau,\lambda) = g(\lambda)$ is independent of $\tau$ and $g\in\mathcal{G}$.
Then $\alpha(\tau+1,\lambda) - \alpha(\tau,\lambda) = 0$, and, by Eq.~\eqref{eq:M_annihilates_G}, $M\alpha(\tau) = Mg = 0$.
Hence a time-independent symmetry transformation acts trivially on the background gauge field, as expected.

\paragraph{Flatness conditions}

The background gauge field obeys the following two types of flatness conditions.
First, we define
\begin{align}
    E\left(\tau+\frac{1}{2}, \hat{\lambda}\right) := A_{s}(\tau+1,\hat{\lambda}) - A_{s}(\tau,\hat{\lambda})
    - \sum_{\lambda\in\Lambda} M_{\hat{\lambda},\lambda}A_{t}\left(\tau+\frac{1}{2}, \lambda\right).
    \label{eq:mixed_field_strength}
\end{align}
A flat gauge field is required to satisfy
\begin{align}
    E\left(\tau+\frac{1}{2}, \hat{\lambda}\right) = 0,
\end{align}
or equivalently
\begin{align}
    A_{s}(\tau+1,\hat{\lambda}) - A_{s}(\tau,\hat{\lambda}) = \sum_{\lambda\in\Lambda} M_{\hat{\lambda},\lambda} A_{t}\left(\tau+\frac{1}{2}, \lambda\right).
    \label{eq:mixed_flatness}
\end{align}
This condition can be understood as follows.
When $A_t=0$, no temporal operators are inserted, and we require the spatial defect configuration to remain unchanged between neighboring time slices, so that $A_s(\tau+1)=A_s(\tau)$.
For nonzero $A_t$, a natural gauge-invariant extension of this requirement is Eq.~\eqref{eq:mixed_flatness}.
This condition states that a change in the spatial defect configuration is accompanied by an insertion of the operator $U(A_t(\tau+\frac{1}{2}))$ between the two time slices.

There can be additional flatness conditions if the chosen set of generators $\hat{\Lambda}$ is locally redundant.
Recalling Eq.~\eqref{eq:hamiltonian_spatial_flatness}, we require that
\begin{align}
    \sum_{\hat{\lambda}\in\hat{\Lambda}} r(\hat{\lambda}) A_{s}(\tau,\hat{\lambda}) = 0 \mod N
    \label{eq:spatial_flatness}
\end{align}
for every time slice $\tau$ and every locally supported $r$ satisfying Eq.~\eqref{eq:local_relation_M}, or equivalently for all $r\in\mathcal{R}$.
If Eq.~\eqref{eq:spatial_flatness} holds at a single time slice $\tau=\tau_0$, then Eq.~\eqref{eq:mixed_flatness} ensures that it holds at every time slice.

Both flatness conditions are invariant under the gauge transformations defined above.

\paragraph{Multiple gauge fields}

When multiple symmetries act on disjoint sets of qudits, possibly with different local Hilbert-space dimensions $N_a$,
we can construct the corresponding background gauge fields separately and combine them into a single background field $A$.
For simplicity, we assume that all qudits have the same local Hilbert-space dimension $N$, unless stated otherwise.
The extension to different $N_a$ is straightforward.

\paragraph{Partition function}
Let us define a partition function in the presence of the background gauge field $A$.
We divide the Euclidean-time circle of circumference $\beta$ into $N_\tau$ intervals of width $\varepsilon=\frac{\beta}{N_{\tau}}$.
We impose periodic boundary conditions in the temporal direction, identifying $\tau=0$ with $\tau=N_\tau$.
We write $A_s^{(k)}:=A_s(k)$ and $A_t^{(k-\frac{1}{2})}:=A_t(k-\frac{1}{2})$.
The transfer matrix for time evolution over an interval of length $t$ is defined by
\begin{align}
    \mathbb{T}_{t}[A_{s}] := e^{-t H[A_{s}]}.
\end{align}
We have
\begin{align}
    U(\alpha) \mathbb{T}_{t}[A_{s}] U^{\dagger}(\alpha) = \mathbb{T}_{t}[A_{s} + M\alpha].
    \label{eq:defect_T_gauge_equivalence}
\end{align}
We define the partition function by
\begin{align}
    Z[\beta; A] = \operatorname{Tr}_{\mathcal{H}_{\mathrm{phys}}}\left[\mathbb{T}_{\varepsilon} [A_{s}^{(N_{\tau})}] U(A_{t}^{(N_{\tau}-\frac{1}{2})})\mathbb{T}_{\varepsilon} [A_{s}^{(N_{\tau}-1)}] U (A_{t}^{(N_{\tau}-\frac{3}{2})}) \cdots \mathbb{T}_{\varepsilon} [A_{s}^{(1)}] U(A_{t}^{(\frac{1}{2})})\right].
\end{align}
Using Eqs.~\eqref{eq:defect_T_gauge_equivalence} and \eqref{eq:mixed_flatness}, we obtain
\begin{align}
    Z[\beta; A] = \operatorname{Tr}_{\mathcal{H}_{\mathrm{phys}}} \left[U(\bar{A}_{t}) \mathbb{T}_{\beta}[A_{s}^{(N_{\tau})}]\right],
\end{align}
where $\bar{A}_{t} = \sum_{\tau=1}^{N_{\tau}} A_{t}^{(\tau-\frac{1}{2})}$.

\subsection{Entanglers for lattice response actions}
\label{sec:entangler}

Here, we construct an entangler that implements the stacking of a phase characterized by a given response action.
For conventional SPT phases described by group cohomology, such a construction was originally developed by Chen et al.~\cite{Chen:2011pg}.
Tsui and Wen~\cite{Tsui:2019ykk} extended this approach to 3+1d SPT phases protected by $\mathbb{Z}_N$ 1-form symmetry, explicitly incorporating background gauge fields.
In the following, we further extend this construction to a broader class of symmetries generated by products of generalized Pauli $X$ operators, including subsystem symmetries.

We assume that the response action is gauge invariant and takes the following form
\begin{align}
    iS_{\Omega}[A] = \sum_{\tau=1}^{N_{\tau}} \Omega[A_{s}^{(\tau+1)},A_{t}^{(\tau+\frac{1}{2})},A_{s}^{(\tau)}]
    \label{eq:form_of_response_action}
\end{align}
for a flat gauge field $A$.
When we write $\Omega[A_{s}^{'},A_{t},A_{s}]$, its arguments always satisfy
\begin{align}
    A_{s}^{'} - A_{s} = MA_{t},\quad
    \sum_{\hat{\lambda}\in\hat{\Lambda}} r(\hat{\lambda}) A_{s}(\hat{\lambda}) = 0,\quad
    \sum_{\hat{\lambda}\in\hat{\Lambda}} r(\hat{\lambda}) A'_{s}(\hat{\lambda}) = 0
\end{align}
for any $r$ satisfying Eq.~\eqref{eq:local_relation_M} with local support, and hence for all $r \in \mathcal{R}$.
We require that
\begin{align}
    \Omega[A_{s},0,A_{s}] = 0
\end{align}
for spatially-flat $A_{s}$.
We also assume that $\Omega$ is local.
All the lattice actions presented in this paper satisfy these assumptions.
Gauge invariance under the transformations in Eq.~\eqref{eq:gauge_transformation_temporal} and Eq.~\eqref{eq:gauge_transformation_spatial} implies
\begin{align}
    \label{eq:gauge_trsf_of_Omega}
    \Omega[A_{s}+MA_{t},A_{t}+s,A_{s}-Ms] + \Omega[A_{s}-Ms,A^{'}_{t}-s,A_{s}-MA^{'}_{t}] &= \Omega[A_{s}+MA_{t},A_{t},A_{s}] + \Omega[A_{s},A^{'}_{t},A_{s}-MA^{'}_{t}],\\
    \label{eq:gauge_trsf_of_Omega2}
    \Omega[A_{s}+MA_{t},A_{t}+c,A_{s}] &= \Omega[A_{s}+MA_{t},A_{t},A_{s}] \quad \text{for }c\in \mathcal{C}.
\end{align}
Rearranging Eq.~\eqref{eq:gauge_trsf_of_Omega}, we find that the following difference depends only on $s$ and $A_{s}$:
\begin{align}
    \begin{split}
    \phi[s,A_{s}] &\coloneqq \Omega[A_{s}-Ms,A'_{t}-s,A_{s}-MA^{'}_{t}] - \Omega[A_{s},A'_{t},A_{s}-MA^{'}_{t}] \\
    &= - (\Omega[A_{s}+MA_{t},A_{t}+s,A_{s}-Ms]-\Omega[A_{s}+MA_{t},A_{t},A_{s}]).
    \end{split}
\end{align}
It satisfies
\begin{align}
    \begin{split}
    &\quad \Omega[A_{s}+MA_{t} - Ms', A_{t} - s' + s, A_{s} - Ms] - \Omega[A_{s}+MA_{t}, A_{t}, A_{s}]\\
    &= (\Omega[A_{s}+MA_{t} - Ms', A_{t} - s' + s, A_{s} - Ms] - \Omega[A_{s}+MA_{t}, A_{t} + s, A_{s} - Ms]) \\
    &\quad - (\Omega[A_{s}+MA_{t}, A_{t}, A_{s}] - \Omega[A_{s}+MA_{t}, A_{t} + s, A_{s} - Ms])\\
    &= \phi[s',A_{s}+MA_{t}] - \phi[s,A_{s}].
    \end{split}
\end{align}
In particular,
\begin{align}
    \Omega[A_{s} - Ms', - s' + s, A_{s} - Ms] = \phi[s',A_{s}] - \phi[s,A_{s}].
    \label{eq:convenient_formula_of_omega_and_phi}
\end{align}
If $g \in \mathcal{G}$, i.e. $Mg = 0$,
\begin{align}
    \phi[s-g,A_{s}] - \phi[s,A_{s}] = \Omega[A_{s} - Ms, g, A_{s} - Ms] = \Omega[A_{s}, g, A_{s}]
    \label{eq:convenient_formula_of_omega_and_phi2}
\end{align}
On the other hand, Eq.~\eqref{eq:gauge_trsf_of_Omega2} implies
\begin{align}
    \phi[s + c,A_{s}] = \phi[s,A_{s}],
    \label{eq:phi_and_local_constraint}
\end{align}
for $c \in \mathcal{C}$.

Applying a gauge transformation to the topological action, we obtain
\begin{align}
    iS_{\Omega}[A] = \sum_{\tau=1}^{N_{\tau}-1} \Omega[A_{s}^{(N_{\tau})},0,A_{s}^{(N_{\tau})}] + \Omega[A_{s}^{(N_{\tau})},\bar{A}_{t},A_{s}^{(N_{\tau})}] = \Omega[A_{s}^{(N_{\tau})},\bar{A}_{t},A_{s}^{(N_{\tau})}] = -\phi[\bar{A}_{t},A_{s}^{(N_{\tau})}],
    \label{eq:gauge_transforming_action}
\end{align}
where $\bar{A}_{t} = \sum_{\tau=1}^{N_{\tau}} A_{t}^{(\tau-\frac{1}{2})}$.
The corresponding partition function is given by
\begin{align}
    Z_{\Omega}[A] = e^{-S_{\Omega}[A]} = e^{i\Omega[A_{s}^{(N_{\tau})},\bar{A}_{t},A_{s}^{(N_{\tau})}]} = e^{-i\phi[\bar{A}_{t},A_{s}^{(N_{\tau})}]}.
\end{align}

Let us construct the entangler $V$ for $S_{\Omega}$.
We define it as
\begin{align}
    V = \sum_{s} e^{i\phi[s,0]} \ket{\{s\}}\bra{\{s\}}.
\end{align}
As shown below, conjugation by $V$ maps symmetric Hamiltonians to symmetric Hamiltonians.
Since $\Omega$ is local, so is $\phi$, which implies that $V$ can be implemented by a finite-depth local unitary circuit.
Consequently, if $H_1$ and $H_2$ belong to the same gapped phase,
$V H_1 V^\dagger$ and $V H_2 V^\dagger$ also belong to the same gapped phase, which may differ from the original one.
Thus, $V$ induces a well-defined map on the set of gapped phases.
We will show that the entangler $V$ implements the stacking of an SPT phase described by Eq.~\eqref{eq:form_of_response_action}.
Eq.~\eqref{eq:phi_and_local_constraint} guarantees that $[V,U(c)]=0$ for $c \in \mathcal{C}$,
and therefore $V$ is consistent with the local constraints.
In particular, we argue that if we begin with the trivial Hamiltonian
\begin{align}
    H_{\mathrm{triv}} = - \sum_{\lambda} \mathcal{X}_{\lambda} + \mathrm{h.c.},
\end{align}
the response theory of the Hamiltonian
\begin{align}
    H' = VH_{\mathrm{triv}}V^{\dagger}
\end{align}
is described by Eq.~\eqref{eq:form_of_response_action} in the low-temperature limit.

To show this, we seek an entangler that maps $H[A_{s}]$ to $H'[A_{s}]$,
\begin{align}
    H'[A_{s}] = V_{A_{s}} H[A_{s}] V^{\dagger}_{A_{s}}.
    \label{eq:action_of_defect_entangler}
\end{align}
We call $V_{A_{s}}$ a defect entangler. In particular, $V_{0} = V$.
We now show that
\begin{align}
    V_{A_{s}} = \sum_{s} e^{i\phi[s,A_{s}]} \ket{\{s\}}\bra{\{s\}}
\end{align}
satisfies Eq.~\eqref{eq:action_of_defect_entangler}.
Like $V$, $V_{A_{s}}$ can be implemented by a finite-depth local unitary circuit.
Eq.~\eqref{eq:phi_and_local_constraint} guarantees that $[V_{A_{s}},U(c)]=0$ for $c\in\mathcal{C}$,
so $V_{A_{s}}$ preserves the physical Hilbert space.

The defect entangler $V_{A_{s}}$ transforms $\mathcal{X}_{\lambda}$ as
\begin{align}
    V_{A_{s}} \mathcal{X}_{\lambda} V^{\dagger}_{A_{s}} = \mathcal{X}_{\lambda} W_{\lambda},
\end{align}
where
\begin{align}
    W_{\lambda}[A_{s}] = \sum_{s} e^{i(\phi[s+\delta_{\lambda}, A_{s}]-\phi[s, A_{s}])} \ket{\{s\}}\bra{\{s\}}.
\end{align}
Using Eq.~\eqref{eq:convenient_formula_of_omega_and_phi}, we obtain
\begin{align}
    W_{\lambda}[A_{s}] = \sum_{s} e^{i\Omega[(A_{s} - Ms) - M\delta_{\lambda}, - \delta_{\lambda}, A_{s} - Ms]} \ket{\{s\}}\bra{\{s\}},
\end{align}
where $\delta_{\lambda}$ is given by
\begin{align}
    \delta_{\lambda}(\lambda') =
    \begin{cases}
        1 & \text{if } \lambda' = \lambda, \\
        0 & \text{otherwise}.
    \end{cases}
\end{align}

$W_{\lambda}[A_{s}]$ has the following properties.
\begin{itemize}
    \item $W_{\lambda}[A_{s}]$ has local support since $V_{A_{s}}$ is a finite-depth local unitary circuit.
    \item $W_{\lambda}[A_{s}]$ is diagonal with respect to the computational basis $\{\ket{s}\}$
    and $\expval{W_{\lambda}[A_{s}]}{s}$ is determined by $\lambda$ and $Ms - A_{s}$,
    implying that $W_{\lambda}[A_{s}]$ can be written in terms of $\omega^{-A_{s}(\hat{\lambda})}\mathcal{Z}^{(\hat{\lambda})}$.
\end{itemize}
On the other hand, the defect entangler $V_{A_{s}}$ commutes with $\mathcal{Z}^{(\hat{\lambda})}$.
Therefore, Eq.~\eqref{eq:action_of_defect_entangler} holds for a symmetric Hamiltonian $H$.
In other words, conjugation by $V_{A_{s}}$ is compatible with the coupling to the background gauge field,
and $V_{A_s}H[A_s]V_{A_s}^\dagger$ gives the defect Hamiltonian associated with $H'=VHV^\dagger$.

If $g \in \mathcal{G}$, i.e. $Mg = 0$, we have
\begin{align}
    U(g) V_{A_{s}} U^{\dagger}(g)
    = \left( \sum_{s} e^{i (\phi[s-g,A_{s}]-\phi[s,A_{s}])} \ket{\{s\}}\bra{\{s\}} \right) V_{A_{s}} = e^{i \Omega[A_{s},g,A_{s}]} V_{A_{s}}.
\end{align}
The second equality follows from Eq.~\eqref{eq:convenient_formula_of_omega_and_phi2}.
Since
\begin{align}
    M\bar{A}_t = \sum_{\tau=1}^{N_\tau}(A_s^{(\tau+1)}-A_s^{(\tau)}) = 0,
\end{align}
the partition function after applying the entangler is given by
\begin{align}
    Z'[\beta; A]
    = \operatorname{Tr}_{\mathcal{H}_{\mathrm{phys}}} \left[U(\bar{A}_{t}) V_{A_{s}^{(N_{\tau})}}\mathbb{T}_{\beta}[A_{s}^{(N_{\tau})}]V^{\dagger}_{A_{s}^{(N_{\tau})}}\right]
    = e^{i \Omega[A_{s}^{(N_{\tau})},\bar{A}_{t},A_{s}^{(N_{\tau})}]} Z[\beta; A] = Z_{\Omega}[A] Z[\beta; A],
\end{align}
which implies that the entangler $V$ implements the stacking of a system described by the response theory $S_{\Omega}$.
Importantly, this identity holds for any $\beta$.

In particular, if we begin with the trivial Hamiltonian $H_{\mathrm{triv}}$,
whose low-temperature partition function is given by $Z[A] \sim 1$ up to a constant independent of $A$,
we obtain the Hamiltonian $H'$, which is described by the response theory $S_{\Omega}$ in the low-temperature limit.

\subsection{Gauge cochain complexes and dual symmetries}
\label{sec:gauge_cochain_complex}

We have obtained the background gauge field and its gauge transformations and flatness conditions entirely from the symmetry action on the Hamiltonian,
as discussed in Section~\ref{sec:construction_of_gauge_fields}.
The temporal component $A_t$ records symmetry insertions between neighboring time slices, while the spatial component $A_s$ specifies the corresponding defect Hamiltonian.
They transform under the gauge transformations given in Eqs.~\eqref{eq:gauge_transformation_temporal} and~\eqref{eq:gauge_transformation_spatial}.
They also satisfy the flatness conditions given in Eqs.~\eqref{eq:mixed_flatness} and~\eqref{eq:spatial_flatness}.
In this subsection, we organize these structures in terms of cochain complexes,
which will be useful for constructing gauge-invariant lattice response theories for SSPT phases in later sections.

\paragraph{Spatial gauge cochain complex}
Here, we introduce a four-term cochain complex called a \emph{spatial gauge cochain complex},
which encodes the spatial structure of the lattice gauge field.
Since $\mathcal{C} \subset \operatorname{Map}(\Lambda,\mathbb{Z}_{N})$, we have an embedding
\begin{align}
    \mathcal{C} \xhookrightarrow{\iota_{\mathcal{C}}} \operatorname{Map}(\Lambda,\mathbb{Z}_{N}).
\end{align}
Similarly, we have an embedding
\begin{align}
    \mathcal{R} \xhookrightarrow{\iota_\mathcal{R}} \operatorname{Hom}(\operatorname{Map}(\hat\Lambda,\mathbb{Z}_{N}),\mathbb{Z}_{N}).
\end{align}
Since every character of a subgroup of a finite abelian group extends to the whole group, the inclusion $\iota_{\mathcal{R}}$ induces a surjective homomorphism
\begin{align}
    p_{\mathcal R}:\operatorname{Map}(\hat\Lambda,\mathbb Z_N)\twoheadrightarrow\operatorname{Hom}(\mathcal R,\mathbb Z_N).
\end{align}
We thus obtain the spatial gauge cochain complex
\begin{align}
    \mathcal{C} \xhookrightarrow{\iota_\mathcal{C}} \operatorname{Map}(\Lambda,\mathbb{Z}_{N}) \xrightarrow{M} \operatorname{Map}(\hat\Lambda,\mathbb{Z}_{N}) \overset{p_{\mathcal{R}}}{\twoheadrightarrow} \operatorname{Hom}(\mathcal{R},\mathbb{Z}_{N}).
    \label{eq:spatial_gauge_cochain_complex}
\end{align}
By construction, we have $M \circ \iota_\mathcal{C} = 0$ and $p_{\mathcal{R}} \circ M = 0$.
The original symmetry actions $\mathcal{G}$ are encoded as
\begin{align}
    \mathcal{G} = \operatorname{Ker} M.
\end{align}

\paragraph{Spacetime gauge cochain complex}

Here, we reformulate the construction of the spacetime lattice gauge field in terms of cochain complexes.
Taking the tensor product of the temporal cochain complex $L_{(1,N)}$ and the spatial gauge cochain complex~\eqref{eq:spatial_gauge_cochain_complex},
we obtain a five-term cochain complex.
We define the \emph{spacetime gauge cochain complex} by retaining its middle three terms:
\begin{align}
    C^{g} \xrightarrow{\dg} C^{gf} \xrightarrow{\df} C^{f},
    \label{eq:spacetime_gauge_cochain_complex}
\end{align}
where
\begin{align}
    C^{g} &= \left(\operatorname{Map}(\Lambda,\mathbb{Z}_{N}) \otimes L_{(1,N)}^{0}\right) \oplus \left(\mathcal{C} \otimes L_{(1,N)}^{1}\right),\\
    C^{gf} &= \left(\operatorname{Map}(\hat\Lambda,\mathbb{Z}_{N}) \otimes L_{(1,N)}^{0}\right) \oplus \left(\operatorname{Map}(\Lambda,\mathbb{Z}_{N}) \otimes L_{(1,N)}^{1}\right),\\
    C^{f} &= \left(\operatorname{Hom}(\mathcal{R},\mathbb{Z}_{N}) \otimes L_{(1,N)}^{0}\right) \oplus \left(\operatorname{Map}(\hat\Lambda,\mathbb{Z}_{N}) \otimes L_{(1,N)}^{1}\right).
\end{align}
Here, the superscripts $g$, $gf$, and $f$ stand for ``gauge transformation,'' ``gauge field,'' and ``field strength (or flatness condition),'' respectively.
A gauge transformation parameter $\xi$, a gauge field configuration $A$, and a field strength $F$ are written as
\begin{align}
    \xi &= (\alpha, c) \in C^{g}, \\
    A &= (A_{s}, A_{t}) \in C^{gf}, \\
    F &= (B, E) \in C^{f}.
\end{align}
The coboundary maps act on these variables as
\begin{align}
    \dg \xi &= (M\alpha, \Delta_{0} \alpha - c),\\
    \df A &= (p_{\mathcal R}(A_s), \Delta_{0}A_{s} - MA_{t}),
\end{align}
where $\Delta_{0}$ denotes the lattice derivative along the temporal direction.
By construction, we have $\df \circ \dg =0$.
The gauge transformation is written as
\begin{align}
    A \to A + \dg \xi,
\end{align}
and the flatness condition is written as
\begin{align}
    \df A = 0.
\end{align}
The gauge-equivalence classes of flat gauge field configurations are classified by
\begin{align}
    H^{gf} := \frac{\operatorname{Ker} \df}{\operatorname{Im} \dg}.
\end{align}

\paragraph{Dual symmetry}
Here, we assume that the symmetry $\mathcal{G}$ is maximally topological.
Taking the dual of Eq.~\eqref{eq:spatial_gauge_cochain_complex} and
using isomorphisms $\operatorname{Map}(\Lambda,\mathbb{Z}_{N}) \simeq \operatorname{Hom}(\operatorname{Map}(\Lambda,\mathbb{Z}_{N}),\mathbb{Z}_{N})$
and $\operatorname{Map}(\hat\Lambda,\mathbb{Z}_{N}) \simeq \operatorname{Hom}(\operatorname{Map}(\hat\Lambda,\mathbb{Z}_{N}),\mathbb{Z}_{N})$,
we obtain
\begin{align}
    \mathcal{R} \rightarrow \operatorname{Map}(\hat\Lambda,\mathbb{Z}_{N}) \xrightarrow{M^T} \operatorname{Map}(\Lambda,\mathbb{Z}_{N}) \rightarrow \operatorname{Hom}(\mathcal{C},\mathbb{Z}_{N}).
    \label{eq:dual_spatial_gauge_cochain_complex}
\end{align}

We consider a new quantum spin system with qudits labeled by $\hat{\Lambda}$, subject to local constraints specified by $\mathcal{R}$.
More precisely,
each qudit labeled by $\hat{\lambda}\in \hat{\Lambda}$ has generalized Pauli operators $\hat{\mathcal{X}}_{\hat{\lambda}},\hat{\mathcal{Z}}_{\hat{\lambda}}$.
The local constraints are given by
\begin{align}
    \prod_{\hat{\lambda}\in \hat{\Lambda}} \hat{\mathcal{X}}_{\hat{\lambda}}^{r(\hat{\lambda})} = 1,
\end{align}
for $r \in \mathcal{R}$.
We consider a symmetry generated by
\begin{align}
    \prod_{\hat{\lambda}\in \hat{\Lambda}} \hat{\mathcal{X}}_{\hat{\lambda}}^{\hat{g}(\hat{\lambda})},\qquad \hat{g} \in \hat{\mathcal{G}}
\end{align}
where $\hat{\mathcal{G}} = \operatorname{Ker} M^{T}$.
The algebra of operators symmetric under $\hat{\mathcal{G}}$ is generated by $\hat{\mathcal{X}}_{\hat{\lambda}}$ and
\begin{align}
    \hat{\mathcal{Z}}^{(\lambda)} = \prod_{\hat{\lambda}\in\hat{\Lambda}}\hat{\mathcal{Z}}_{\hat{\lambda}}^{M_{\hat{\lambda},\lambda}},\qquad \lambda\in\Lambda.
\end{align}
Indeed, the exponents of symmetric $\mathcal{Z}$-type operators are precisely the elements of $\hat{\mathcal{G}}^\perp = (\operatorname{Ker}M^T)^\perp = \operatorname{Im}M$.
Since $\mathcal{G}$ is maximally topological, $\mathcal{C}$ includes every $c\in \operatorname{Map}(\Lambda,\mathbb{Z}_{N})$ with local support satisfying
\begin{align}
    \prod_{\lambda\in\Lambda} \left(\hat{\mathcal{Z}}^{(\lambda)}\right)^{c(\lambda)} = 1,
\end{align}
implying that $\mathcal{C}$ encodes all local flatness conditions associated with the symmetry $\hat{\mathcal{G}}$.
Therefore, we identify Eq.~\eqref{eq:dual_spatial_gauge_cochain_complex} as a spatial gauge cochain complex associated with the symmetry $\hat{\mathcal{G}}$.
We refer to $\hat{\mathcal{G}}$ as the dual symmetry of $\mathcal{G}$.
Since $\mathcal{R}$ is generated by all locally supported elements of $\operatorname{Ker}M^T$, the dual symmetry $\hat{\mathcal{G}}$ is also maximally topological.
By considering a tensor product with $L_{(1,N)}$, we obtain a spacetime gauge cochain complex and therefore a dual gauge field.

We call a symmetry \emph{self-dual} if its spatial gauge cochain complex is isomorphic to that of its dual symmetry, with the isomorphism preserving the local structure.

As we will see in Sec.~\ref{sec:intersection_pairing_general},
an intersection pairing can be defined between gauge fields associated with two dual symmetries, $\mathcal{G}$ and $\hat{\mathcal{G}}$.
Although we will not demonstrate this explicitly, this pairing can serve as a kernel for gauging.
In particular, gauging $\mathcal{G}$ gives rise to the dual symmetry $\hat{\mathcal{G}}$.

For example, in $d$ spatial dimensions, a conventional $\mathbb{Z}_N$ $p$-form symmetry is dual to a $\mathbb{Z}_N$ $(d-p-1)$-form symmetry~\cite{Gaiotto:2014kfa}.
Thus, it is self-dual when $2p+1=d$.
A 2+1d linear subsystem symmetry, which will be discussed in Sec.~\ref{sec:2+1d_linear_SSPT}, is also self-dual~\cite{Cao:2023doz}.

\paragraph{Higher and lower degrees}

We can extend the gauge cochain complex to higher and lower degrees.
For example, in 3+1d, a background gauge field for a $\mathbb{Z}_N$ 1-form symmetry is described by a 2-cochain.
Its gauge transformation is parametrized by a 1-cochain, while its field strength is a 3-cochain.
These structures are sufficient to describe flat gauge fields and their gauge equivalence classes.
However, we can further extend the cochain complex to include 0-cochains and 4-cochains, which encode redundancies in gauge transformation parameters and Bianchi identities, respectively.
We now explain how such extensions arise in our construction.

We first discuss the extension to lower degrees.
We choose a set of local generators of $\mathcal{C}$ labeled by $\Lambda_{\mathcal{C}}$.
We assume that every locally supported element $c\in\mathcal{C}$ can be expressed as a linear combination of these generators whose supports lie within a bounded neighborhood of the support of $c$, independently of the system size.
We denote these generators by $c_{\lambda_{\mathcal{C}}}\in\mathcal{C}$ for $\lambda_{\mathcal{C}}\in\Lambda_{\mathcal{C}}$.
However, these generators may satisfy further local relations of the form
\begin{align}
    \sum_{\lambda_{\mathcal{C}} \in \Lambda_{\mathcal{C}}} \ell(\lambda_{\mathcal{C}}) c_{\lambda_{\mathcal{C}}} = 0,
\end{align}
for some $\ell(\lambda_{\mathcal{C}})$ with a local support.
We denote by $\mathcal{L}$ the subgroup of $\operatorname{Map}(\Lambda_{\mathcal{C}},\mathbb{Z}_{N})$ generated by all such locally supported relations.
Then, we have
\begin{align}
    \mathcal{L} \xhookrightarrow{\iota_\mathcal{L}} \operatorname{Map}(\Lambda_{\mathcal{C}},\mathbb{Z}_{N}) \rightarrow \operatorname{Map}(\Lambda,\mathbb{Z}_{N}) \xrightarrow{M} \operatorname{Map}(\hat\Lambda,\mathbb{Z}_{N}) \overset{p_{\mathcal{R}}}{\twoheadrightarrow} \operatorname{Hom}(\mathcal{R},\mathbb{Z}_{N}).
\end{align}
The composition of any two consecutive coboundary maps vanishes.
We can repeat this procedure by introducing a set of locally supported generators of $\mathcal{L}$ and considering further local relations among them.

Similarly, we can extend the complex to higher degrees.
We choose a set of local generators of $\mathcal{R}$ labeled by $\Lambda_{\mathcal{R}}$, subject to the same locality condition imposed on the generators of $\mathcal{C}$.
These generators define a surjective homomorphism
\begin{align}
    \rho_{\mathcal{R}}:\operatorname{Map}(\Lambda_{\mathcal{R}},\mathbb{Z}_N)\twoheadrightarrow\mathcal{R}.
\end{align}
We define $\mathcal{K}$ as the subgroup of $\operatorname{Ker}\rho_{\mathcal{R}}$ generated by its locally supported elements.
Then we have an inclusion map $\mathcal{K} \hookrightarrow \operatorname{Map}(\Lambda_{\mathcal{R}},\mathbb{Z}_{N})$.
Using an isomorphism $\operatorname{Map}(\Lambda_{\mathcal{R}},\mathbb{Z}_{N}) \simeq \operatorname{Hom}(\operatorname{Map}(\Lambda_{\mathcal{R}},\mathbb{Z}_{N}),\mathbb{Z}_{N})$,
we obtain
\begin{align}
    \mathcal{C} \xhookrightarrow{\iota_\mathcal{C}} \operatorname{Map}(\Lambda,\mathbb{Z}_{N}) \xrightarrow{M} \operatorname{Map}(\hat\Lambda,\mathbb{Z}_{N}) \rightarrow \operatorname{Map}(\Lambda_{\mathcal{R}},\mathbb{Z}_{N}) \overset{p_{\mathcal{K}}}{\twoheadrightarrow} \operatorname{Hom}(\mathcal{K},\mathbb{Z}_{N}).
\end{align}
The composition of any two consecutive coboundary maps vanishes.
We can repeat the same procedure to obtain higher degrees.

In this way, we can extend the spatial gauge cochain complex to higher and lower degrees.
Taking its tensor product with $L_{(1,N)}$, we obtain a corresponding extension of the spacetime gauge cochain complex.

Although these extensions provide additional information about relations among gauge transformations and flatness conditions,
the three-term spacetime gauge cochain complex is sufficient to describe flat gauge field configurations and their gauge equivalence classes.
Nevertheless, such extensions will be useful for constructing a gauge-invariant response action for 3+1d planar subsystem symmetries with two foliations, as we will see in Sec.~\ref{sec:typeII_SSPT}.

\section{SSPT phases from intersection pairings}
\label{sec:intersection_pairing}

In this section, we construct a class of SSPT phases whose response partition functions take the form
\begin{align}
    Z[A,\hat{A}] = \exp\left(\frac{2\pi i}{N}\langle \hat{A},A\rangle\right),
\end{align}
where $\langle \hat{A},A\rangle$ denotes a bilinear intersection pairing between two mutually dual $\mathbb{Z}_N$ gauge fields $A$ and $\hat{A}$.
We show that this construction reproduces various SSPT phases studied in previous literature.

\subsection{General construction}
\label{sec:intersection_pairing_general}

We begin with the Hamiltonian formalism.
Following the construction in Sec.~\ref{sec:construction_of_gauge_fields},
we consider a qudit system with qudits labeled by elements of $\Lambda$, local constraints $\mathcal{C}$, and a symmetry group $\mathcal{G}\supset\mathcal{C}$.
In this subsection, we assume that $\mathcal{G}$ is maximally topological.
We choose a proper generating set $\hat{\Lambda}$ of local symmetric $\mathcal{Z}$-type operators, which determines a matrix $M_{\hat{\lambda},\lambda}$.
We denote the gauge field associated with $\mathcal{G}$ by $A$.

We introduce a second qudit system carrying the dual symmetry defined in Sec.~\ref{sec:gauge_cochain_complex}.
Its qudits are labeled by elements of $\hat{\Lambda}$, with generalized Pauli operators $\hat{\mathcal{X}}_{\hat{\lambda}}$ and $\hat{\mathcal{Z}}_{\hat{\lambda}}$.
The local constraints are specified by $\mathcal{R}$, and the symmetry group is $\hat{\mathcal{G}}=\operatorname{Ker}M^T$.
We denote the corresponding gauge field by $\hat{A}$.

The gauge transformations of $A$ and $\hat{A}$ are given by
\begin{align}
    A_{t}\left(\tau+\frac{1}{2},\lambda\right) \longrightarrow\;& A_{t}\left(\tau+\frac{1}{2},\lambda\right) + \alpha(\tau+1,\lambda) - \alpha(\tau,\lambda) - c_{\tau+\frac{1}{2}}(\lambda), \\
    A_{s}(\tau,\hat{\lambda}) \longrightarrow\;& A_{s}(\tau,\hat{\lambda}) + \sum_{\lambda\in\Lambda} M_{\hat{\lambda},\lambda} \alpha(\tau,\lambda), \\
    \hat{A}_{t}\left(\tau+\frac{1}{2},\hat\lambda\right) \longrightarrow\;& \hat{A}_{t}\left(\tau+\frac{1}{2},\hat\lambda\right) + \hat{\alpha}(\tau+1,\hat\lambda) - \hat{\alpha}(\tau,\hat\lambda) - r_{\tau+\frac{1}{2}}(\hat\lambda), \\
    \hat{A}_{s}(\tau,\lambda) \longrightarrow\;& \hat{A}_{s}(\tau,\lambda) + \sum_{\hat\lambda\in\hat\Lambda} M_{\hat{\lambda},\lambda} \hat{\alpha}(\tau,\hat\lambda),
\end{align}
for $c_{\tau+\frac{1}{2}} \in \mathcal{C}$ and $r_{\tau+\frac{1}{2}} \in \mathcal{R}$.
The flatness conditions are
\begin{align}
    A_{s}(\tau+1,\hat{\lambda}) - A_{s}(\tau,\hat{\lambda}) &= \sum_{\lambda\in\Lambda} M_{\hat{\lambda},\lambda} A_{t} \left(\tau+\frac{1}{2},\lambda\right), \\
    \sum_{\hat{\lambda}\in\hat{\Lambda}} r(\hat{\lambda}) A_{s}(\tau,\hat{\lambda}) &= 0, \quad \text{for all }r\in\mathcal{R},\\
    \hat{A}_{s}(\tau+1,\lambda) - \hat{A}_{s}(\tau,\lambda) &= \sum_{\hat\lambda\in\hat\Lambda} M_{\hat{\lambda},\lambda} \hat{A}_{t} \left(\tau+\frac{1}{2},\hat\lambda\right), \\
    \sum_{\lambda\in\Lambda} c(\lambda) \hat{A}_{s}(\tau,\lambda) &= 0, \quad \text{for all }c\in\mathcal{C}.
\end{align}
Here and below, all equalities involving $\mathbb{Z}_N$-valued quantities are understood modulo $N$.

We can then construct the following intersection pairing:
\begin{align}
    \begin{split}
    \langle\hat{A},A\rangle &= \sum_{\tau} \left\{\left(\hat{A}_{s}(\tau)\right)^T A_{t}\left(\tau+\frac{1}{2}\right) + \left(\hat{A}_{t}\left(\tau-\frac{1}{2}\right)\right)^T A_{s}(\tau)\right\}\\
    &= \sum_{\tau} \left\{\sum_{\lambda \in \Lambda} \hat{A}_{s}(\tau,\lambda) A_{t}\left(\tau+\frac{1}{2},\lambda\right) + \sum_{\hat\lambda \in \hat\Lambda} \hat{A}_{t}\left(\tau-\frac{1}{2},\hat{\lambda}\right) A_{s}(\tau,\hat{\lambda})\right\}.
    \end{split}
\end{align}
This pairing is invariant under the gauge transformations of both $A$ and $\hat{A}$ on a spacetime with periodic temporal boundary conditions,
provided that the gauge fields satisfy the flatness conditions.
It therefore defines a gauge-invariant response partition function
\begin{align}
    Z[A,\hat{A}] = \exp\left(\frac{2\pi i}{N}\langle\hat{A},A\rangle\right).
    \label{eq:intersection_partition_function}
\end{align}

We show that the partition function~\eqref{eq:intersection_partition_function} is identically unity for all flat background gauge fields
if and only if $\mathcal{G}=\mathcal{C}$ and $\hat{\mathcal{G}}=\mathcal{R}$.
As in Eq.~\eqref{eq:gauge_transforming_action},
we can use gauge transformations along the temporal circle
to represent any flat background by time-independent spatial components $f\in\mathcal{R}^\perp$ and $\hat{f}\in\mathcal{C}^\perp$,
together with temporal holonomies $g\in\mathcal{G}$ and $\hat{g}\in\hat{\mathcal{G}}$.
The partition function then takes the form
\begin{align}
    Z[A,\hat{A}] = \exp\left(\frac{2\pi i}{N} \left(\hat{f}^{T} g + \hat{g}^{T} f \right)\right).
\end{align}
Since these holonomies and spatial components can be chosen independently,
the partition function is identically unity if and only if
$\hat{\mathcal{G}} \subset (\mathcal{R}^\perp)^{\perp},\,\mathcal{G} \subset (\mathcal{C}^\perp)^\perp$.
Using $(\mathcal{R}^\perp)^\perp=\mathcal{R}$ and $(\mathcal{C}^\perp)^\perp=\mathcal{C}$,
these conditions are equivalent to $\hat{\mathcal{G}} \subset \mathcal{R},\, \mathcal{G} \subset \mathcal{C}$.
Together with $\mathcal{R}\subset\hat{\mathcal{G}}$ and $\mathcal{C}\subset\mathcal{G}$, this proves the claim.
This criterion guarantees the nontriviality of the response actions constructed in the following subsections.

Following Sec.~\ref{sec:entangler}, the corresponding entangler can be constructed explicitly as
\begin{align}
    V = \sum_{s,\hat{s}}\exp\left(\frac{2\pi i}{N}\hat{s}^T M s\right) \ket{s,\hat{s}}\bra{s,\hat{s}},
\end{align}
where $s\in\operatorname{Map}(\Lambda,\mathbb{Z}_N)$ and $\hat{s}\in\operatorname{Map}(\hat{\Lambda},\mathbb{Z}_N)$ label the computational basis states of the two qudit systems.
By construction, $V$ commutes with both symmetry groups $\mathcal{G}$ and $\hat{\mathcal{G}}$.
The entangler $V$ is a generalized cluster-state entangler associated with the weighted bipartite graph specified by $M$~\cite{Looi:2008dzc}.
Various cluster-type SSPT models have been constructed in previous studies~\cite{You:2018oai,Devakul:2018fhz,Devakul:2019noj,Zhang:2023ldq,ParayilMana:2024txy,Okuda:2024azp,Furukawa:2025flp,ParayilMana:2025nxw}.

We now interpret this construction in terms of gauge cochain complexes.
The spatial gauge cochain complexes associated with $\mathcal{G}$ and $\hat{\mathcal{G}}$,
given in Eqs.~\eqref{eq:spatial_gauge_cochain_complex} and~\eqref{eq:dual_spatial_gauge_cochain_complex},
are dual to each other by construction.
Since the temporal cochain complex $L_{(1,N)}$ is self-dual up to a half-lattice translation,
the corresponding spacetime gauge cochain complexes are also dual to each other.
This duality induces a bilinear pairing between their cohomology groups,
which is explicitly realized by the intersection pairing constructed above.

\subsection{2+1d linear SSPT}
\label{sec:2+1d_linear_SSPT}

In this subsection, we illustrate our construction using 2+1d SSPT phases protected by linear subsystem symmetries.
In particular, we reproduce the lattice models of Ref.~\cite{You:2018oai} using entanglers derived from the corresponding response actions.
See also Refs.~\cite{Burnell:2021reh,Cao:2023rrb,Jia:2025eff} for related discussions.

\paragraph{Symmetry action}
We consider a square lattice $\Lambda=\mathbb{Z}_{L_x}\times\mathbb{Z}_{L_y}$ with periodic boundary conditions and a $\mathbb{Z}_{N}$ qudit at each site.
The linear subsystem symmetry is generated by
\begin{align}
    U_{x}(x_0) = \prod_{y\in\mathbb{Z}_{L_y}} \mathcal{X}_{(x_0,y)},\qquad U_{y}(y_0) = \prod_{x\in\mathbb{Z}_{L_x}} \mathcal{X}_{(x,y_0)}.
\end{align}
These symmetry operators satisfy the global relation
\begin{align}
    \prod_{x\in\mathbb{Z}_{L_x}} U_{x}(x) = \prod_{y\in\mathbb{Z}_{L_y}} U_{y}(y).
\end{align}
The algebra of local symmetric operators is generated by $\mathcal{X}_{(x,y)}$ and the plaquette operators
\begin{align}
    \mathcal{Z}_{(x+1,y+1)}\mathcal{Z}^{\dagger}_{(x+1,y)}\mathcal{Z}^{\dagger}_{(x,y+1)}\mathcal{Z}_{(x,y)}.
\end{align}

\paragraph{Gauge fields}
Applying the method in Sec.~\ref{sec:construction_of_gauge_fields}, we obtain a lattice gauge field associated with the symmetry.
See Refs.~\cite{Cao:2023rrb,Jia:2025eff} for related discussions in terms of symmetry topological field theory (SymTFT).
Background lattice gauge fields for linear subsystem symmetries were previously formulated in Refs.~\cite{Cao:2022lig,Cao:2023doz}.

We now consider two species of qudits, labeled by $A$ and $B$, with local Hilbert-space dimensions $N_A$ and $N_B$, respectively.
Each species carries independent linear subsystem symmetries, with a $\mathbb{Z}_{N_A}$ or $\mathbb{Z}_{N_B}$ symmetry acting along each row and column.
We denote the corresponding background gauge fields by $A=(A_0,A_{12})$ and $B=(B_0,B_{12})$.

The gauge transformations are given by
\begin{align}
    A_{0} &\to A_{0} + \Delta_{0}k,\\
    A_{12} &\to A_{12} + \Delta_{1}\Delta_{2} k,\\
    B_{0} &\to B_{0} + \Delta_{0}k',\\
    B_{12} &\to B_{12} + \Delta_{1}\Delta_{2} k'.
\end{align}
The flatness conditions for the gauge fields are given by
\begin{align}
    \Delta_{0}A_{12} - \Delta_{1}\Delta_{2}A_{0} &= 0,\\
    \Delta_{0}B_{12} - \Delta_{1}\Delta_{2}B_{0} &= 0.
\end{align}
The gauge transformations and flatness conditions are encoded
in the spacetime gauge cochain complex given in Eq.~\eqref{eq:spacetime_gauge_cochain_complex}.

\paragraph{Lattice response action}

We now construct a lattice response action using the intersection pairing introduced in Sec.~\ref{sec:intersection_pairing}.
For each $I=A,B$, the spacetime gauge cochain complex is given by the tensor product $L_{(1,N_I)}\otimes L_{(2,N_I)}$.
Since both factors are self-dual, the spacetime gauge cochain complex is also self-dual, allowing us to construct self-intersection pairings for the gauge fields $A$ and $B$.
Furthermore, the mixed intersection pairing between the $\mathbb{Z}_{N_A}$ and $\mathbb{Z}_{N_B}$ gauge fields is defined modulo $\gcd(N_A,N_B)$.
The resulting gauge-invariant response action is
\begin{align}
    \begin{split}
    -S_{k_{A},k_{B},k_{AB}} = &\frac{2\pi i k_{A}}{N_{A}} \sum_{n} (A_{0}(n)A_{12}(n+\hat{0})+A_{12}(n)A_{0}(n+\hat{1}+\hat{2}))\\
    +& \frac{2\pi i k_{B}}{N_{B}} \sum_{n} (B_{0}(n)B_{12}(n+\hat{0})+B_{12}(n)B_{0}(n+\hat{1}+\hat{2}))\\
    +& \frac{2\pi i k_{AB}}{\gcd(N_{A},N_{B})} \sum_{n} (A_{0}(n)B_{12}(n+\hat{0})+A_{12}(n)B_{0}(n+\hat{1}+\hat{2})).
    \end{split}
    \label{eq:linear_SSPT_action}
\end{align}

Exchanging the order of $A$ and $B$ in the mixed intersection pairing is equivalent to a relative spacetime translation of the gauge fields.
More generally, translating one of the gauge fields by a finite spatial vector does not affect the strong SSPT class,
although it may change the response by stacking with weak SSPT phases.
See Ref.~\cite{Devakul:2018fhz} for the definition of strong and weak SSPT phases and Appendix~\ref{app:weak_SSPT} for further details.

The integer labels are identified as $k_{A}\sim k_{A} + N_{A}, k_{B} \sim k_{B} + N_{B}, k_{AB} \sim k_{AB} + \gcd(N_{A},N_{B})$.
This is consistent with the classification
$\mathbb{Z}_{N_{A}}\times \mathbb{Z}_{N_{B}} \times \mathbb{Z}_{\gcd(N_{A},N_{B})}$
of strong SSPT phases~\cite{Devakul:2018fhz}.

\paragraph{Lattice Hamiltonian}

We now construct lattice Hamiltonians realizing these SSPT phases using the entangler introduced in Sec.~\ref{sec:entangler}.
For the response action above, the entangler takes the form
\begin{align}
    V = \sum_{s_A,s_B} e^{i\phi[(s_A,s_B),0]}
    \ket{\{s_A\},\{s_B\}}\bra{\{s_A\},\{s_B\}},
\end{align}
where $s_A\in\operatorname{Map}(\Lambda,\mathbb{Z}_{N_A})$ and $s_B\in\operatorname{Map}(\Lambda,\mathbb{Z}_{N_B})$.
Using Eq.~\eqref{eq:convenient_formula_of_omega_and_phi}, we obtain
\begin{align}
    \begin{split}
    \phi[(s_A,s_B),0] ={}&
    \frac{2\pi k_A}{N_A} \sum_{r\in\Lambda} s_A(r)\Delta_1\Delta_2 s_A(r)
    + \frac{2\pi k_B}{N_B} \sum_{r\in\Lambda} s_B(r) \Delta_1\Delta_2 s_B(r)\\
    &+ \frac{2\pi k_{AB}}{\gcd(N_A,N_B)}
    \sum_{r\in\Lambda}s_A(r)\Delta_1\Delta_2s_B(r).
    \end{split}
\end{align}

We apply the entangler to the trivial paramagnetic Hamiltonian to obtain an SSPT Hamiltonian.
For simplicity, we focus on $N_A=N_B=2$.
To make the lattice structure more transparent, we depict the $B$ qubits at plaquette centers by shifting their positions by $(-1/2,-1/2)$.

\begin{itemize}
    \item If $k_{AB} = 1, k_{A} = k_{B} = 0$, the resulting Hamiltonian is
    \begin{align}
        H = - \sum_{A} \input{figures/double-lin-sspt.tikz} -\sum_{B} \input{figures/double-lin-sspt-sononi.tikz}.
    \end{align}
    \item If $k_{A} = 1, k_{B} = k_{AB} = 0$, the resulting Hamiltonian is
    \begin{align}
        \label{eq:single-lin-sspt}
        H = +\sum_{A} \input{figures/single-lin-sspt.tikz} - \sum_{B} \mathcal{X}_{B}.
    \end{align}
\end{itemize}
These Hamiltonians reproduce the lattice models of Refs.~\cite{You:2018oai,Devakul:2018fhz}.\footnote{The sign of the first term in Eq.~\eqref{eq:single-lin-sspt} differs from that in Ref.~\cite{You:2018oai}. For even $L_x$ and $L_y$, this difference can be removed by a symmetric finite-depth local unitary circuit $\prod_{r\in\Lambda}\mathcal{Z}_A(r)$.}

\paragraph{Continuum description}

We introduce scalar fields $\Phi_A$ and $\Phi_B$ imposing the flatness conditions of $A$ and $B$.
They are subject to the identifications
\begin{align}
    \Phi_{I} \sim \Phi_{I} + 2\pi (m^I_1(x_1)+m^I_2(x_2))\qquad I = A,B,
\end{align}
where $m_1^I(x_1)$ and $m_2^I(x_2)$ are integer-valued piecewise constant functions.

Motivated by the lattice construction, a formal continuum counterpart of the lattice response action is given by
\begin{align}
    \begin{split}
    -S={}&\frac{iN_{A}}{2\pi} \int \Phi_{A} (\partial_{0}A_{12}-\partial_{1}\partial_{2}A_{0}) + \frac{iN_{B}}{2\pi} \int \Phi_{B} (\partial_{0}B_{12}-\partial_{1}\partial_{2}B_{0})\\
    &+ \frac{ik_{A}N_{A}}{\pi} \int A_{0}A_{12} + \frac{ik_{B}N_{B}}{\pi} \int B_{0}B_{12} + \frac{ik_{AB}\mathrm{lcm}(N_{A},N_{B})}{2\pi} \int (A_{0}B_{12}+A_{12}B_{0}),
    \end{split}
\end{align}
where the gauge transformations are given by
\begin{align}
    &A_{0} \to A_{0} + \partial_{0} \alpha_{A}, \qquad A_{12} \to A_{12} + \partial_{1}\partial_{2} \alpha_{A},\\
    &B_{0} \to B_{0} + \partial_{0} \alpha_{B}, \qquad B_{12} \to B_{12} + \partial_{1}\partial_{2} \alpha_{B},\\
    &\Phi_{A} \to \Phi_{A} + 2k_{A}\alpha_{A} + \frac{k_{AB}N_{B}}{\gcd(N_{A},N_{B})} \alpha_{B}, \\
    &\Phi_{B} \to \Phi_{B} + \frac{k_{AB}N_{A}}{\gcd(N_{A},N_{B})} \alpha_{A} + 2k_{B}\alpha_{B}.
\end{align}
The gauge transformation parameters themselves have identifications
\begin{align}
    \alpha_{I} \sim \alpha_{I} + 2\pi (n^I_1(x_1)+n^I_2(x_2))\qquad I = A,B,
\end{align}
where $n^I_1(x_1)$ and $n^I_2(x_2)$ are integer-valued piecewise constant functions.
For $k_B = k_{AB} = 0$, the action is closely related to the continuum SSPT theory studied in Ref.~\cite{Burnell:2021reh},
although the coefficient of the $A_0 A_{12}$ term differs.
A related subtlety concerning the consistency of lattice SSPT response actions was discussed in Ref.~\cite{Cao:2023rrb}.

\paragraph{Generalization}
The construction extends to an arbitrary finite Abelian onsite symmetry group $\prod_{\alpha=1}^{m}\mathbb{Z}_{N_\alpha}$.
For each factor, we introduce a gauge field $A^{(\alpha)}=(A_0^{(\alpha)},A_{12}^{(\alpha)})$ with the gauge transformations
\begin{align}
    A_{0}^{(\alpha)} &\to A_{0}^{(\alpha)} + \Delta_{0}k^{(\alpha)},\\
    A_{12}^{(\alpha)} &\to A_{12}^{(\alpha)} + \Delta_{1}\Delta_{2}k^{(\alpha)},
\end{align}
and the flatness condition
\begin{align}
    \Delta_{0}A^{(\alpha)}_{12} - \Delta_{1}\Delta_{2}A^{(\alpha)}_{0} = 0.
\end{align}
The topological action is given by
\begin{align}
    \begin{split}
    -S =& \sum_{\alpha} \frac{2\pi i k_{\alpha}}{N_{\alpha}} \sum_{n}(A^{(\alpha)}_{0}(n)A^{(\alpha)}_{12}(n+\hat{0})+A^{(\alpha)}_{12}(n)A^{(\alpha)}_{0}(n+\hat{1}+\hat{2}))\\
    &+ \sum_{\alpha<\beta}\frac{2\pi i k_{\alpha\beta}}{\gcd{(N_{\alpha},N_{\beta})}} \sum_{n}(A^{(\alpha)}_{0}(n)A^{(\beta)}_{12}(n+\hat{0})+A^{(\alpha)}_{12}(n)A^{(\beta)}_{0}(n+\hat{1}+\hat{2})).
    \end{split}
\end{align}
The independent coefficients $k_\alpha$ and $k_{\alpha\beta}$ reproduce the strong SSPT classification $\prod_{\alpha=1}^{m}\mathbb{Z}_{N_\alpha}\times\prod_{\alpha<\beta}\mathbb{Z}_{\gcd(N_\alpha,N_\beta)}$~\cite{Devakul:2018fhz}.

\subsection{4+1d SSPT for the X-cube model}
\label{sec:SSPT_for_X_cube}
In this subsection, we study a family of 4+1d SSPT phases, including one whose anomaly inflow cancels the mixed 't Hooft anomaly of the 3+1d $\mathbb{Z}_N$ X-cube model.
A continuum description of the SSPT response and its anomaly inflow was developed in Ref.~\cite{Burnell:2021reh}.
For $N=2$, Ref.~\cite{Okuda:2024azp} constructed both a lattice response theory and a foliated cluster-state model realizing the corresponding bulk-boundary anomaly inflow.
Although the latter formulation is closely related to the continuum tensor gauge theory, an explicit component-wise dictionary between the continuum tensor gauge fields and the lattice cochains has not been established.
Moreover, the lattice construction of Ref.~\cite{Okuda:2024azp} was formulated for $\mathbb{Z}_2$ degrees of freedom.

Here, we establish the connection between the two formulations and extend the lattice construction to general $\mathbb{Z}_N$.
We first provide a lattice counterpart of the continuum SSPT response of Ref.~\cite{Burnell:2021reh},
giving an explicit component-wise correspondence between the continuum and lattice fields.
From the lattice response theory, we derive a finite-depth entangler for the SSPT phases,
whose $N=2$ case reproduces the foliated cluster-state construction of Ref.~\cite{Okuda:2024azp}.

\paragraph{Symmetry action}
Throughout this subsection, $(ijk)$ denotes the cyclic permutations $(123)$, $(231)$, and $(312)$.
We consider a four-dimensional hypercubic lattice with $\mathbb{Z}_N$ qudits on two sets of lattice cells,
\begin{align}
    \Lambda = \left(\bigsqcup_{(ijk)} \Lambda_{ij}\right) \sqcup \Lambda_4,\qquad
    \hat{\Lambda} = \left(\bigsqcup_{(ijk)} \hat{\Lambda}^{ij}\right) \sqcup \left(\bigsqcup_{(ijk)} \hat{\Lambda}_{4}^{[jk]i}\right),
\end{align}
where $\Lambda_{ij}$, $\Lambda_4$, $\hat{\Lambda}^{ij}$, and $\hat{\Lambda}_{4}^{[jk]i}$
denote the sets of $ij$-plaquettes, $4$-links, $k$-links with a label $(i,j,k)=(1,2,3),(2,3,1),(3,1,2)$,
and $4$-links with a label $i=1,2,3$, respectively.
We denote their generalized Pauli operators by
$\mathcal{X}_{ij}(n)$, $\mathcal{Z}_{ij}(n)$, $\mathcal{X}_4(n)$, $\mathcal{Z}_4(n)$,
$\hat{\mathcal{X}}^{ij}(n)$, $\hat{\mathcal{Z}}^{ij}(n)$,
$\hat{\mathcal{X}}_{4}^{[jk]i}(n)$, and $\hat{\mathcal{Z}}_{4}^{[jk]i}(n)$, respectively.

We consider two types of $\mathbb{Z}_{N}$ subsystem symmetries.
The first, denoted by $\mathbb{Z}_{N}^{(L)}$, is generated by
\begin{align}
    U(g) = \prod_{n}\left((\mathcal{X}_4(n))^{g_4(n)} \prod_{(ijk)} (\mathcal{X}_{ij}(n))^{g_{ij}(n)}\right)
\end{align}
such that
\begin{align}
    \Delta_{4} g_{ij} - \Delta_{i}\Delta_{j} g_{4} = 0,\\
    \Delta_{i} g_{jk} - \Delta_{j} g_{ki} = 0.
\end{align}
We impose local constraints $U(\delta \kappa) = 1$ on the physical Hilbert space where $(\delta \kappa)_{ij} = \Delta_{i}\Delta_{j}\kappa, (\delta \kappa)_{4} = \Delta_{4}\kappa$.
The algebra of local $\mathbb{Z}_{N}^{(L)}$-symmetric operators is generated by $\mathcal{X}_{ij}(n)$, $\mathcal{X}_4(n)$, and
\begin{align}
    \mathcal{Z}_{4ij}(n) &= \mathcal{Z}_{ij}(n+\hat{4})\mathcal{Z}^{\dagger}_{ij}(n) \mathcal{Z}^{\dagger}_{4}(n+\hat{i}+\hat{j}) \mathcal{Z}_{4}(n+\hat{i}) \mathcal{Z}_{4}(n+\hat{j}) \mathcal{Z}^{\dagger}_{4}(n), \\
    \mathcal{Z}_{[ij]k}(n) &= \mathcal{Z}_{jk}^{\dagger}(n+\hat{i})\mathcal{Z}_{jk}(n) \mathcal{Z}_{ki}(n+\hat{j})\mathcal{Z}^{\dagger}_{ki}(n).
\end{align}

The dual subsystem symmetry, denoted by $\mathbb{Z}_{N}^{(P)}$, is generated by
\begin{align}
    \hat{U}(\hat{g}) = \prod_{(ijk)}\prod_{n} \left((\hat{\mathcal{X}}^{ij}(n))^{\hat{g}^{ij}(n)}  (\hat{\mathcal{X}}_{4}^{[jk]i}(n))^{\hat{g}_4^{[jk]i}(n)} \right)
\end{align}
such that
\begin{align}
    \sum_{(ijk)} \Delta_{i}\Delta_{j}\hat{g}^{ij} = 0,\\
    \Delta_{4}\hat{g}^{ij} - \Delta_{k}(\hat{g}_{4}^{[ki]j}-\hat{g}_{4}^{[jk]i}) = 0.
\end{align}
We impose local constraints $\hat{U}(\delta \hat{\kappa}) = 1$ on the physical Hilbert space where $(\delta \hat{\kappa})^{ij} = \Delta_{k}(\hat{\kappa}^{[ki]j}-\hat{\kappa}^{[jk]i}), (\delta \hat{\kappa})_{4}^{[jk]i} = \Delta_{4}\hat{\kappa}^{[jk]i}$.
We also impose the local constraints
\begin{align}
    \prod_{(ijk)} \hat{\mathcal{X}}_{4}^{[jk]i}(n) = 1.
    \label{eq:superlocal_constraint}
\end{align}
The algebra of local $\mathbb{Z}_{N}^{(P)}$-symmetric operators is generated by $\hat{\mathcal{X}}^{ij}(n)$, $\hat{\mathcal{X}}_{4}^{[jk]i}(n)$,
\begin{align}
    \hat{\mathcal{Z}}_{\hat C}(n) &= \prod_{(ijk)} \left(\hat{\mathcal{Z}}^{ij}(n+\hat{i}+\hat{j}) (\hat{\mathcal{Z}}^{ij}(n+\hat{i}))^\dagger (\hat{\mathcal{Z}}^{ij}(n+\hat{j}))^\dagger \hat{\mathcal{Z}}^{ij}(n)\right),\\
    \hat{\mathcal{Z}}_{4}^{ij}(n) &= \hat{\mathcal{Z}}^{ij}(n+\hat{4})(\hat{\mathcal{Z}}^{ij}(n))^\dagger (\hat{\mathcal{Z}}_{4}^{[ki]j}(n+\hat{k}))^{\dagger} \hat{\mathcal{Z}}_{4}^{[ki]j}(n)\hat{\mathcal{Z}}_{4}^{[jk]i}(n+\hat{k}) (\hat{\mathcal{Z}}_{4}^{[jk]i}(n))^\dagger.
\end{align}

\paragraph{Gauge fields}

Applying the general construction in Sec.~\ref{sec:construction_of_gauge_fields},
we obtain the following gauge transformations and flatness conditions.
We choose the signs of the gauge field components to facilitate comparison with Ref.~\cite{Burnell:2021reh}.

The gauge field for $\mathbb{Z}_{N}^{(L)}$ subsystem symmetry has components $C_{0ij},C_{4ij},C_{[04]},C_{[ij]k}$
with the relations
\begin{align}
    C_{0ij} = C_{0ji},\quad C_{4ij} = C_{4ji},\quad C_{[ij]k} = - C_{[ji]k}.
\end{align}
Gauge transformations are given by
\begin{align}
    C_{0ij} &\to C_{0ij}+\Delta_{0} \alpha_{ij} - \Delta_{i}\Delta_{j}\alpha_{0},\\
    C_{4ij} &\to C_{4ij}+\Delta_{4} \alpha_{ij} - \Delta_{i}\Delta_{j}\alpha_{4},\\
    C_{[04]} &\to C_{[04]} + \Delta_{0}\alpha_{4} - \Delta_{4}\alpha_{0}, \\
    C_{[ij]k} &\to C_{[ij]k} - \Delta_{i}\alpha_{jk} + \Delta_{j}\alpha_{ki},
\end{align}
where $\alpha_{ij} = \alpha_{ji}$.
They satisfy the following flatness conditions
\begin{align}
    \Delta_{4}C_{[ij]k} + \Delta_{i} C_{4jk} - \Delta_{j} C_{4ik} &= 0,\\
    \Delta_{0}C_{[ij]k} + \Delta_{i} C_{0jk} - \Delta_{j} C_{0ik} &= 0,\\
    \Delta_{i}\Delta_{j} C_{[04]} + \Delta_{0}C_{4ij} - \Delta_{4}C_{0ij} &= 0,\\
    C_{[12]3} + C_{[23]1} + C_{[31]2} &= 0.
\end{align}

The gauge field for $\mathbb{Z}_{N}^{(P)}$ subsystem symmetry has the components $\hat{C}_{0}^{ij},\hat{C}_{[04]}^{[ij]k},\hat{C},\hat{C}_{4}^{ij}$
with the relations
\begin{align}
    \hat{C}_{0}^{ij} = \hat{C}_{0}^{ji},\quad \hat{C}_{4}^{ij} = \hat{C}_{4}^{ji},\quad\hat{C}_{[04]}^{[ij]k} = - \hat{C}_{[04]}^{[ji]k}.
\end{align}
Gauge transformations are given by
\begin{align}
    \hat{C}_{0}^{ij} &\to \hat{C}_{0}^{ij} + \Delta_{0} \hat{\alpha}^{ij} - \Delta_{k} (\hat{\alpha}_{0}^{[ki]j}-\hat{\alpha}_{0}^{[jk]i}), \\
    \hat{C}_{[04]}^{[ij]k} &\to \hat{C}_{[04]}^{[ij]k} + \Delta_{0}\hat{\alpha}_{4}^{[ij]k} - \Delta_{4}\hat{\alpha}_{0}^{[ij]k} - \hat{c}, \\
    \hat{C} &\to \hat{C} + \sum_{(ijk)} \Delta_{i}\Delta_{j}\hat{\alpha}^{ij},\\
    \hat{C}_{4}^{ij} &\to \hat{C}_{4}^{ij} + \Delta_{4}\hat{\alpha}^{ij} - \Delta_{k} (\hat{\alpha}_{4}^{[ki]j}-\hat{\alpha}_{4}^{[jk]i}).
\end{align}
The additional gauge transformation parametrized by $\hat{c}$ originates from the local constraint~\eqref{eq:superlocal_constraint}.
It acts as a common shift of the three components of $\hat{C}_{[04]}^{[ij]k}$.
The corresponding tensor components in the notation of Ref.~\cite{Burnell:2021reh} are given by
\begin{align}
    \hat{C}_{[04]}^{k(ij)} = \hat{C}_{[04]}^{[ki]j} - \hat{C}_{[04]}^{[jk]i}.
\end{align}
These components are invariant under the common shift and satisfy the relation
\begin{align}
    \sum_{(ijk)}\hat{C}_{[04]}^{k(ij)} = 0.
\end{align}
Thus, the two descriptions have the same number of independent components.

They satisfy the following flatness conditions
\begin{align}
    \Delta_{4}\hat{C} - \sum_{(ijk)} \Delta_{i}\Delta_{j} \hat{C}_{4}^{ij} &= 0,\\
    \Delta_{0}\hat{C} - \sum_{(ijk)} \Delta_{i}\Delta_{j} \hat{C}_{0}^{ij} &= 0,\\
    \Delta_{k}(\hat{C}_{[04]}^{[ki]j}-\hat{C}_{[04]}^{[jk]i}) + \Delta_{0}\hat{C}_{4}^{ij} - \Delta_{4}\hat{C}_{0}^{ij} &= 0.
\end{align}

$\mathbb{Z}_{N}^{(L)}$ and $\mathbb{Z}_{N}^{(P)}$ symmetries are both maximally topological.
They are dual to each other in the sense explained in Sec.~\ref{sec:gauge_cochain_complex}.

\paragraph{Lattice response action}
Applying the intersection pairing constructed in Sec.~\ref{sec:intersection_pairing_general},
we obtain the following action
\begin{align}
    \begin{split}
    -S_p = \frac{2\pi i p}{N}\sum_{n}& \left\{\sum_{(ijk)} C_{0ij}(n)\hat{C}_{4}^{ij}(n+\hat{0}-\hat{k}) - \sum_{(ijk)} C_{4ij}(n)\hat{C}_{0}^{ij}(n-\hat{k}+\hat{4})\right. \\
    &+ \left. C_{[04]}(n)\hat{C}(n+\hat{0}-\hat{1}-\hat{2}-\hat{3}+\hat{4}) - \sum_{(ijk)} C_{[ij]k}(n)\hat{C}_{[04]}^{[ij]k}(n)\right\},
    \end{split}
    \label{eq:xcube_intersection}
\end{align}
where $p \in \mathbb{Z}_{N}$.
This is a lattice version of the response action of the SSPT phase for the $\mathbb{Z}_{N}^{(L)}\times \mathbb{Z}_{N}^{(P)}$ subsystem symmetry discussed in \cite{Burnell:2021reh}.

\paragraph{Entangler}
From the lattice response action, we obtain the corresponding entangler
\begin{align}
    \begin{split}
    V_{p} &= \sum_{s_{ij},s_4,\hat{s}^{ij},\hat{s}_{4}^{[jk]i}}\exp\Bigg[
    \frac{2\pi ip}{N} \sum_n \sum_{(ijk)}\Bigg(
      s_{ij}(n)\left\{
        \Delta_4 \hat{s}^{ij}(n-\hat k) - \Delta_{k} (\hat{s}_{4}^{[ki]j}-\hat{s}_{4}^{[jk]i})(n-\hat k)
      \right\}
      \\
      &\qquad\qquad\qquad
      +s_4(n) \Delta_i \Delta_j \hat{s}^{ij}(n - \hat 1 - \hat 2 - \hat 3 + \hat 4)
    \Bigg)
  \Bigg] \ket{\{s_{ij}\},\{s_4\},\{\hat{s}^{ij}\},\{\hat{s}_{4}^{[jk]i}\}}\bra{\{s_{ij}\},\{s_4\},\{\hat{s}^{ij}\},\{\hat{s}_{4}^{[jk]i}\}},
    \end{split}
\end{align}
which coincides with the cluster entangler considered in Ref.~\cite{Okuda:2024azp} when $N=2$ and $p=1$.
Eq.~\eqref{eq:xcube_intersection} also agrees with the intersection pairing derived in Ref.~\cite{Okuda:2024azp}.

\paragraph{Continuum description}
We now compare our lattice response theory with the continuum description of Ref.~\cite{Burnell:2021reh}.
A formal continuum counterpart of our lattice response action is given by
\begin{align}
    \begin{split}
    &\quad -S_p\\
    &= \frac{iN}{2\pi}\int \sum_{(ijk)}\Big\{
    \hat{B}_0^{[ij]k}(\partial_4 C_{[ij]k} + \partial_i C_{4jk} - \partial_j C_{4ik})
    - \hat{B}_4^{[ij]k}(\partial_0 C_{[ij]k} + \partial_i C_{0jk} - \partial_j C_{0ik})
    - \hat{B}^{ij}(\partial_i\partial_j C_{[04]} + \partial_0 C_{4ij} - \partial_4 C_{0ij})
    \Big\}\\
    &\quad+ \frac{iN}{2\pi}\int \left\{ B_0\left(\partial_4\hat{C} - \sum_{(ijk)}\partial_i\partial_j\hat{C}_4^{ij}\right)
    - B_4\left(\partial_0\hat{C} - \sum_{(ijk)}\partial_i\partial_j\hat{C}_0^{ij}\right)
     - \sum_{(ijk)} B_{ij}\Big\{
    \partial_k(\hat{C}_{[04]}^{[ki]j} - \hat{C}_{[04]}^{[jk]i})
    + \partial_0\hat{C}_4^{ij} - \partial_4\hat{C}_0^{ij}
    \Big\}\right\} \\
    &\quad+ \frac{ipN}{2\pi} \int \left\{\sum_{(ijk)} C_{0ij}\hat{C}_{4}^{ij} - \sum_{(ijk)} C_{4ij}\hat{C}_{0}^{ij} + C_{[04]}\hat{C} - \sum_{(ijk)} C_{[ij]k}\hat{C}_{[04]}^{[ij]k}\right\}.
    \end{split}
    \label{eq:sspt_for_xcube_continuum}
\end{align}
Here, $(\hat{B}_0^{[ij]k},\hat{B}_4^{[ij]k},\hat{B}^{ij})$ and $(B_0,B_4,B_{ij})$ are dynamical tensor gauge fields that impose the flatness conditions on $C$ and $\hat{C}$, respectively.
The background gauge fields satisfy $C_{[12]3} + C_{[23]1} + C_{[31]2} = 0$.
The gauge field components are identified with those of Ref.~\cite{Burnell:2021reh} by replacing the continuum coordinates $\tau$ and $w$ with $0$ and $4$, respectively.
We choose the sign conventions of the dynamical tensor gauge fields $B$ and $\hat{B}$ to facilitate comparison with Ref.~\cite{Burnell:2021reh}.
Under the background gauge transformations, the gauge fields transform as
\begin{align}
    C_{0ij} &\to C_{0ij}+\partial_{0} \alpha_{ij} - \partial_{i}\partial_{j}\alpha_{0},\\
    C_{4ij} &\to C_{4ij}+\partial_{4} \alpha_{ij} - \partial_{i}\partial_{j}\alpha_{4},\\
    C_{[04]} &\to C_{[04]} + \partial_{0}\alpha_{4} - \partial_{4}\alpha_{0}, \\
    C_{[ij]k} &\to C_{[ij]k} - \partial_{i}\alpha_{jk} + \partial_{j}\alpha_{ki}, \\
    \hat{C}_{0}^{ij} &\to \hat{C}_{0}^{ij} + \partial_{0} \hat{\alpha}^{ij} - \partial_{k} (\hat{\alpha}_{0}^{[ki]j}-\hat{\alpha}_{0}^{[jk]i}), \\
    \hat{C}_{[04]}^{[ij]k} &\to \hat{C}_{[04]}^{[ij]k} + \partial_{0}\hat{\alpha}_{4}^{[ij]k} - \partial_{4}\hat{\alpha}_{0}^{[ij]k} - \hat{c}, \\
    \hat{C} &\to \hat{C} + \sum_{(ijk)} \partial_{i}\partial_{j}\hat{\alpha}^{ij},\\
    \hat{C}_{4}^{ij} &\to \hat{C}_{4}^{ij} + \partial_{4}\hat{\alpha}^{ij} - \partial_{k} (\hat{\alpha}_{4}^{[ki]j}-\hat{\alpha}_{4}^{[jk]i}),\\
    B_0 &\to B_{0} - p\alpha_{0},\\
    B_4 &\to B_{4} - p\alpha_{4},\\
    B_{ij} &\to B_{ij} - p\alpha_{ij},\\
    \hat{B}_0^{[ij]k} &\to \hat{B}_0^{[ij]k} + p\hat{\alpha}_0^{[ij]k},\\
    \hat{B}_4^{[ij]k} &\to \hat{B}_4^{[ij]k} + p\hat{\alpha}_4^{[ij]k},\\
    \hat{B}^{ij} &\to \hat{B}^{ij} + p\hat{\alpha}^{ij}.
\end{align}
For $p = 1$, the continuum expression agrees with the Euclidean Lagrangian of Ref.~\cite{Burnell:2021reh}, up to our convention for the sign of the action.
The dynamical tensor gauge fields also possess independent gauge transformations, and the gauge parameters are subject to gauge-for-gauge redundancies, as discussed in Ref.~\cite{Burnell:2021reh}.

\subsection{2-foliated SSPT in 3+1d}
\label{sec:Shimamura_2fol}

In this subsection, we discuss a 3+1d 2-foliated SSPT phase whose continuum tensor gauge theory description and anomaly inflow were studied in Ref.~\cite{Shimamura:2024kwf}.

\paragraph{Symmetry action}
We consider a three-dimensional cubic lattice with $\mathbb{Z}_N$ qudits on two sets of lattice cells,
\begin{align}
    \Lambda = \Lambda_{12} \sqcup \Lambda_3,\qquad \hat{\Lambda} = \{\text{sites}\},
\end{align}
where $\Lambda_{12}$ and $\Lambda_3$ denote the sets of $12$-plaquettes and $3$-links, respectively.
We label each site, link, and plaquette by its base point $n$,
and denote the generalized Pauli operators by
$\mathcal{X}_{12}(n)$, $\mathcal{Z}_{12}(n)$, $\mathcal{X}_3(n)$, $\mathcal{Z}_3(n)$,
$\hat{\mathcal{X}}(n)$, and $\hat{\mathcal{Z}}(n)$.

We consider two types of $\mathbb{Z}_N$ subsystem symmetries.
The first, denoted by $\mathbb{Z}_N^C$, acts on the qudits in $\Lambda$ through
\begin{align}
    U(g) = \prod_n \mathcal{X}_{12}(n)^{g_{12}(n)}\mathcal{X}_3(n)^{g_3(n)},
\end{align}
where $g = (g_{12},g_3)$ satisfies
\begin{align}
    \Delta_3 g_{12} - \Delta_1 \Delta_2 g_3 = 0.
\end{align}
We impose the local constraints $U(D\kappa) = 1$ on the physical Hilbert space, where
\begin{align}
    (D\kappa)_{12} = \Delta_1 \Delta_2 \kappa,\qquad (D\kappa)_3 = \Delta_3 \kappa.
\end{align}
The algebra of local $\mathbb{Z}_N^C$-symmetric operators is generated by $\mathcal{X}_{12}(n)$, $\mathcal{X}_3(n)$, and
\begin{align}
    \mathcal{Z}_{312}(n) := \mathcal{Z}_{12}(n+\hat{3})\mathcal{Z}_{12}^{-1}(n)
    \mathcal{Z}_3^{-1}(n)\mathcal{Z}_3(n+\hat{1})\mathcal{Z}_3(n+\hat{2})\mathcal{Z}_3^{-1}(n+\hat{1}+\hat{2}).
\end{align}

The second subsystem symmetry, denoted by $\hat{\mathbb{Z}}_N^{\hat C}$, acts on the qudits in $\hat{\Lambda}$ through
\begin{align}
    \hat U(\hat g) = \prod_n \hat{\mathcal{X}}(n)^{\hat g(n)},
\end{align}
where $\hat g$ satisfies
\begin{align}
    \Delta_1 \Delta_2 \hat g = 0,\qquad \Delta_3 \hat g = 0.
\end{align}
Equivalently, the symmetry is generated by the planar operators
\begin{align}
    \hat U_x(x_0) = \prod_{y,z} \hat{\mathcal{X}}(x_0,y,z),\qquad \hat U_y(y_0) = \prod_{x,z} \hat{\mathcal{X}}(x,y_0,z).
\end{align}
These generators satisfy the relation
\begin{align}
    \prod_{x_0} \hat U_x(x_0) = \prod_{y_0} \hat U_y(y_0).
\end{align}
We refer to this symmetry as a \emph{2-foliated planar subsystem symmetry}, which will also appear in later sections.
The algebra of local $\hat{\mathbb{Z}}_N^{\hat C}$-symmetric operators is generated by $\hat{\mathcal{X}}(n)$ and
\begin{align}
    \hat{\mathcal{Z}}_{12}(n) := \hat{\mathcal{Z}}(n)\hat{\mathcal{Z}}^{-1}(n+\hat{1})\hat{\mathcal{Z}}^{-1}(n+\hat{2})\hat{\mathcal{Z}}(n+\hat{1}+\hat{2}),\qquad
    \hat{\mathcal{Z}}_3(n) := \hat{\mathcal{Z}}(n+\hat{3})\hat{\mathcal{Z}}^{-1}(n).
\end{align}

$\mathbb{Z}_{N}^{C}$ and $\hat{\mathbb{Z}}_{N}^{\hat{C}}$ symmetries are both maximally topological.
They are dual to each other in the sense explained in Sec.~\ref{sec:gauge_cochain_complex}.

\paragraph{Gauge fields}

Applying the general construction in Sec.~\ref{sec:construction_of_gauge_fields},
we obtain the following gauge transformations and flatness conditions.
We choose the signs of the gauge field components to facilitate comparison with Ref.~\cite{Shimamura:2024kwf}.

The gauge field for $\mathbb{Z}_{N}^{C}$ is described by $C = (C_{012}, C_{312}, C_{[03]})$.
The gauge transformations are given by
\begin{align}
    C_{012} &\to C_{012} + \Delta_{0} \alpha_{12} - \Delta_{1}\Delta_{2} \alpha_{0},\\
    C_{312} &\to C_{312} + \Delta_{3} \alpha_{12} - \Delta_{1}\Delta_{2} \alpha_{3},\\
    C_{[03]} &\to C_{[03]} + \Delta_{0} \alpha_{3} - \Delta_{3}\alpha_{0}.
\end{align}
The flatness condition for the $\mathbb{Z}_{N}^{C}$ gauge field is
\begin{align}
    \Delta_3 C_{012} - \Delta_0 C_{312} - \Delta_1\Delta_2 C_{[03]} = 0.
\end{align}

The gauge field for $\hat{\mathbb{Z}}_{N}^{\hat{C}}$ is described by $(\hat{C}, \hat{C}_{0}^{12}, \hat{C}_{3}^{12})$.
The gauge transformations are given by
\begin{align}
    \hat{C} &\to \hat{C} + \Delta_{1}\Delta_{2} \hat{\alpha},\\
    \hat{C}_{0}^{12} &\to \hat{C}_{0}^{12} + \Delta_{0} \hat{\alpha},\\
    \hat{C}_{3}^{12} &\to \hat{C}_{3}^{12} + \Delta_{3} \hat{\alpha}.
\end{align}
The flatness conditions for the $\hat{\mathbb{Z}}_{N}^{\hat{C}}$ gauge field are given by
\begin{align}
    \Delta_{0}\hat{C} - \Delta_{1}\Delta_{2}\hat{C}_{0}^{12} &= 0, \\
    \Delta_{3}\hat{C} - \Delta_{1}\Delta_{2}\hat{C}_{3}^{12} &= 0, \\
    \Delta_{0} \hat{C}_{3}^{12} - \Delta_{3} \hat{C}_{0}^{12} &= 0.
\end{align}

\paragraph{Lattice response action}

Applying the intersection pairing constructed in Sec.~\ref{sec:intersection_pairing_general},
we obtain the following lattice response action:
\begin{align}
    -S_{p} = \frac{2\pi i p}{N}\sum_{n}\left\{\hat{C}_{0}^{12}(n) C_{312}(n+\hat{0}) - \hat{C}_{3}^{12}(n) C_{012}(n+\hat{3}) - \hat{C}(n) C_{[03]}(n+\hat{1}+\hat{2})\right\},
    \label{eq:2-foliated_intersection}
\end{align}
where $p\in \mathbb{Z}_{N}$.

\paragraph{Entangler}

We obtain the entangler
\begin{align}
    V = \sum_{s_{12},s_{3},\hat{s}}e^{\frac{2\pi i p}{N}\sum_{n} \hat{s}(n) (\Delta_{3} s_{12}(n) - \Delta_{1}\Delta_{2} s_{3}(n))}
    \ket{\{s_{12}\},\{s_{3}\},\{\hat{s}\}}\bra{\{s_{12}\},\{s_{3}\},\{\hat{s}\}}.
\end{align}
For $N = 2$ and $p = 1$, this entangler reproduces the foliated cluster-state entangler associated with the 2+1d plaquette Ising model constructed in
Ref.~\cite{Okuda:2024azp}, up to a relabeling of the plaquette and vertical-link qubits and lattice translations.
Eq.~\eqref{eq:2-foliated_intersection} also agrees with the intersection pairing derived in Ref.~\cite{Okuda:2024azp}.

\paragraph{Continuum description}

The lattice construction motivates the following formal continuum description:
\begin{align}
    - S &= \frac{iN}{2\pi} \int \hat{\beta}^{12}(\partial_3 C_{012} - \partial_0 C_{312} - \partial_1\partial_2 C_{[03]}) \nonumber\\
    &\quad + \frac{iN}{2\pi} \int \Big[
        \beta_0(\partial_3\hat{C} - \partial_1\partial_2\hat{C}_3^{12})
        + \beta_{12}(\partial_3\hat{C}_0^{12} - \partial_0\hat{C}_3^{12})
        - \beta_3(\partial_0\hat{C} - \partial_1\partial_2\hat{C}_0^{12})
    \Big] \nonumber\\
    &\quad - \frac{iNp}{2\pi} \int (C_{012}\hat{C}_3^{12} - C_{312}\hat{C}_0^{12} + C_{[03]}\hat{C}).
\end{align}
Here, $\hat{\beta}^{12}$ and $(\beta_0,\beta_{12},\beta_3)$ are auxiliary fields imposing the flatness conditions on the background gauge fields.

The gauge transformations are given by
\begin{align}
    C_{012} &\to C_{012} + \partial_0\Lambda_{12} - \partial_1\partial_2\Lambda_0,\\
    C_{312} &\to C_{312} + \partial_3\Lambda_{12} - \partial_1\partial_2\Lambda_3,\\
    C_{[03]} &\to C_{[03]} + \partial_0\Lambda_3 - \partial_3\Lambda_0, \\
    \hat{C}_0^{12} &\to \hat{C}_0^{12} + \partial_0\hat{\Lambda}^{12},\\
    \hat{C} &\to \hat{C} + \partial_1\partial_2\hat{\Lambda}^{12},\\
    \hat{C}_3^{12} &\to \hat{C}_3^{12} + \partial_3\hat{\Lambda}^{12},\\
    \hat{\beta}^{12} &\to \hat{\beta}^{12} - p \hat{\Lambda}^{12} + 2\pi\hat{s}^1 - 2\pi\hat{s}^2,\\
    \beta_0 &\to \beta_0 + p\Lambda_0 + \partial_0 s,\\
    \beta_{12} &\to \beta_{12} + p\Lambda_{12} + \partial_1\partial_2 s,\\
    \beta_3 &\to \beta_3 + p\Lambda_3 + \partial_3 s.
\end{align}
Here, $\hat{s}^k$ is an integer-valued function of $x_k$ for $k = 1,2$, and $s$ is a gauge transformation parameter.
For $p = -1$, this continuum theory reduces to that of Ref.~\cite{Shimamura:2024kwf},
including its Boltzmann weight convention.

\section{SSPT phases beyond intersection pairing}
\label{sec:beyond_intersection}

\subsection{2-foliated Bockstein SSPT in 3+1d}
\label{sec:typeII_SSPT}

We next study 3+1d SSPT phases protected by 2-foliated planar subsystem symmetries.
Devakul et al.~\cite{Devakul:2019duj} classified strong planar SSPT phases with Abelian fracton duals and constructed explicit lattice models for them.

The response theories considered here are subsystem analogues of the 2+1d type-I and type-II SPT phases protected by finite Abelian 0-form symmetries.
For example, a type-II SPT phase protected by two $\mathbb{Z}_N$ 0-form symmetries has a response action of the form
\begin{align}
    \frac{2\pi i p}{N} \int A \cup \Bock(B),
    \label{eq:usual_TypeII_action}
\end{align}
where $A$ and $B$ are $\mathbb{Z}_N$ gauge fields and $\Bock$ denotes the Bockstein homomorphism associated with the short exact sequence
\begin{align}
    0 \longrightarrow \mathbb{Z}_N \xrightarrow{\times N} \mathbb{Z}_{N^2} \xrightarrow{\operatorname{mod} N} \mathbb{Z}_N \longrightarrow 0.
\end{align}

\paragraph{Symmetry action}
We consider a three-dimensional cubic lattice with $M$ species of qudits on sites, labeled by $I = 1,\ldots,M$.
The $I$-th species has $\mathbb{Z}_{N_I}$ qudits with generalized Pauli operators $\mathcal{X}^{(I)}_{(x_1,x_2,x_3)}$ and $\mathcal{Z}^{(I)}_{(x_1,x_2,x_3)}$.

For each species, the planar subsystem symmetry is generated by
\begin{align}
    U_{1}^{(I)}(x_1) &= \prod_{x_2,x_3} \mathcal{X}^{(I)}_{(x_1,x_2,x_3)},\\
    U_{2}^{(I)}(x_2) &= \prod_{x_1,x_3} \mathcal{X}^{(I)}_{(x_1,x_2,x_3)}.
\end{align}
These generators satisfy the global relation
\begin{align}
    \prod_{x_1} U_{1}^{(I)}(x_1) = \prod_{x_2} U_{2}^{(I)}(x_2).
\end{align}
For each $I$, this symmetry has the same structure as the $\mathbb{Z}_{N_I}^{\hat{C}}$ subsystem symmetry discussed in Sec.~\ref{sec:Shimamura_2fol}.

\paragraph{Gauge cochain complex}
Following the systematic construction in Sec.~\ref{sec:construction_of_gauge_fields}, we can introduce a background gauge field for each species.
Its gauge transformations and flatness conditions are the same as those of the gauge field associated with the $\mathbb{Z}_{N_I}^{\hat{C}}$ symmetry discussed in Sec.~\ref{sec:Shimamura_2fol}.

Here, we reformulate this structure using the gauge cochain complex introduced in Sec.~\ref{sec:gauge_cochain_complex}.
The three-term spacetime gauge cochain complex encoding the gauge transformations and flatness conditions can be extended to a four-term complex,
which will be useful for constructing Bockstein response actions.

We consider a four-dimensional hypercubic spacetime lattice.
Let $\mathcal{K}_i$ denote the following collections of lattice cells:
\begin{align}
    \mathcal{K}_0 &= \{\text{sites}\},\\
    \mathcal{K}_1 &= \{\text{$0$-links, $3$-links, $12$-plaquettes}\},\\
    \mathcal{K}_2 &= \{\text{$03$-plaquettes, $012$-cubes, $123$-cubes}\},\\
    \mathcal{K}_3 &= \{\text{$0123$-hypercubes}\}.
\end{align}
We define the cochain groups by
\begin{align}
    C_{\text{2-fol}}^i(\mathbb{Z}_N) = \operatorname{Map}(\mathcal{K}_i,\mathbb{Z}_N).
\end{align}
In the following, we suppress the coefficient group $\mathbb{Z}_N$ when no confusion can arise.

The four-term gauge cochain complex is
\begin{align}
    C_{\text{2-fol}}^0 \xrightarrow{\delta_0} C_{\text{2-fol}}^1 \xrightarrow{\delta_1} C_{\text{2-fol}}^2 \xrightarrow{\delta_2} C_{\text{2-fol}}^3.
    \label{eq:extended_gauge_cochain_complex}
\end{align}
For $\alpha \in C_{\text{2-fol}}^0$, $A \in C_{\text{2-fol}}^1$, and $B \in C_{\text{2-fol}}^2$, the coboundary operators are defined by
\begin{align}
    (\delta_0 \alpha)_0 &= \Delta_0 \alpha,\quad (\delta_0 \alpha)_3 = \Delta_3 \alpha,\quad (\delta_0 \alpha)_{12} = \Delta_1\Delta_2 \alpha,\\
    (\delta_1 A)_{03} &= \Delta_0 A_3 - \Delta_3 A_0,\\
    (\delta_1 A)_{012} &= \Delta_0 A_{12} - \Delta_1\Delta_2 A_0,\\
    (\delta_1 A)_{123} &= \Delta_3 A_{12} - \Delta_1\Delta_2 A_3,\\
    (\delta_2 B)_{0123} &= \Delta_3 B_{012} - \Delta_0 B_{123} - \Delta_1\Delta_2 B_{03},
\end{align}
where $\Delta_\mu$ denotes the forward difference in the $\mu$-direction.
These operators satisfy
\begin{align}
    \delta_1 \delta_0 = 0,\qquad \delta_2 \delta_1 = 0.
\end{align}

The background gauge field $A \in C_{\text{2-fol}}^1$ associated with the 2-foliated planar subsystem symmetry transforms as
\begin{align}
    A \to A + \delta_0 \alpha,
\end{align}
and satisfies the flatness condition
\begin{align}
    \delta_1 A = 0.
\end{align}

Interestingly, a cochain $B \in C_{\text{2-fol}}^2$ can be interpreted as a background gauge field associated with
the same type of symmetry as the $\mathbb{Z}_N^C$ subsystem symmetry discussed in Sec.~\ref{sec:Shimamura_2fol}.
After relabeling its components, the gauge transformation and flatness condition take the form
\begin{align}
    B \to B + \delta_1 \beta,\qquad \delta_2 B = 0,
\end{align}
where $\beta \in C_{\text{2-fol}}^1$.

As discussed in Sec.~\ref{sec:Shimamura_2fol}, these two types of gauge fields admit a gauge-invariant intersection pairing on a closed spacetime lattice,
\begin{align}
    \langle A,B\rangle_{N} = \sum_n \bigl\{ A_0(n) B_{123}(n+\hat{0}) - A_3(n) B_{012}(n+\hat{3}) - A_{12}(n) B_{03}(n+\hat{1}+\hat{2}) \bigr\} \mod N.
    \label{eq:2fol_typeII_pairing}
\end{align}
The pairing is invariant under the above gauge transformations when $\delta_1 A = 0$ and $\delta_2 B = 0$.
This gauge invariance follows from the self-duality of the complex~\eqref{eq:extended_gauge_cochain_complex}, which is constructed as the tensor product of one copy of $L_{(2,N)}$ and two copies of $L_{(1,N)}$.

\paragraph{Bockstein response action}
We now use the extended gauge cochain complex to construct Bockstein response actions for 2-foliated planar subsystem symmetries.

We introduce the Bockstein homomorphism
\begin{align}
    \Bock_{N,K}: H_{\text{2-fol}}^1(\mathbb{Z}_N) \to H_{\text{2-fol}}^2(\mathbb{Z}_K)
\end{align}
associated with the short exact sequence
\begin{align}
    0 \longrightarrow \mathbb{Z}_K \xrightarrow{\times N} \mathbb{Z}_{NK} \xrightarrow{\operatorname{mod} N} \mathbb{Z}_N \longrightarrow 0.
\end{align}
Equivalently, it can be obtained by reducing modulo $K$ the integral Bockstein homomorphism associated with
\begin{align}
    0 \longrightarrow \mathbb{Z} \xrightarrow{\times N} \mathbb{Z} \xrightarrow{\operatorname{mod} N} \mathbb{Z}_N \longrightarrow 0.
\end{align}

For a flat gauge field $A \in C_{\text{2-fol}}^1(\mathbb{Z}_N)$, let $\tilde{A}$ be an integer lift of $A$.
Since $\delta_1 A = 0 \mod N$, the integer cochain $\delta_1 \tilde{A}$ is divisible by $N$.
We therefore define
\begin{align}
    \Bock_{N,K}(A) = \frac{\delta_1 \tilde{A}}{N} \mod K.
    \label{eq:2fol_Bockstein}
\end{align}
Its components are explicitly given by
\begin{align}
    (\Bock_{N,K}(A))_{03}(n) &= \frac{\Delta_0 \tilde{A}_3(n) - \Delta_3 \tilde{A}_0(n)}{N} \mod K,\\
    (\Bock_{N,K}(A))_{012}(n) &= \frac{\Delta_0 \tilde{A}_{12}(n) - \Delta_1\Delta_2 \tilde{A}_0(n)}{N} \mod K,\\
    (\Bock_{N,K}(A))_{123}(n) &= \frac{\Delta_3 \tilde{A}_{12}(n) - \Delta_1\Delta_2 \tilde{A}_3(n)}{N} \mod K.
\end{align}

Using the intersection pairing~\eqref{eq:2fol_typeII_pairing}, we construct gauge-invariant response actions by pairing one background gauge field with the Bockstein of another.
The resulting action is
\begin{align}
    -S = \sum_{I<J} \frac{2\pi i k_{IJ}}{N_{I}} \langle A^{I},\Bock_{N_{J},N_{I}}(A^{J})\rangle_{N_{I}}
    + \sum_{I} \frac{2\pi i k_{II}}{N_{I}} \langle A^{I},\Bock_{N_{I},N_{I}}(A^{I})\rangle_{N_{I}},
    \label{eq:typeII_response_action}
\end{align}
where $k_{IJ}, k_{II} \in \mathbb{Z}_{N_I}$.

We first determine the periodicity of the coefficients $k_{IJ}$ for $I < J$.
By definition, $N_J \Bock_{N_J,N_I}(A^{J}) = \delta_1\tilde{A}^{J} \mod N_I$, which is exact.
Since $N_I \Bock_{N_J,N_I}(A^{J}) = 0$ as a cochain, it follows that $\gcd(N_I,N_J)\Bock_{N_J,N_I}(A^{J})$ is also exact.
Consequently, the mixed response action is invariant under
\begin{align}
    k_{IJ} \to k_{IJ} + \gcd(N_I,N_J).
\end{align}

For flat background gauge fields on a closed spacetime lattice, the phase
\begin{align}
    \exp\left\{2\pi i\left(\frac{1}{N_I}\langle A^{I},\Bock_{N_J,N_I}(A^{J})\rangle_{N_I} +\frac{1}{N_J}\langle A^{J},\Bock_{N_I,N_J}(A^{I})\rangle_{N_J}\right) \right\}
\end{align}
represents a weak SSPT response,
as can be shown by an argument analogous to that in Appendix~\ref{app:weak_SSPT} for 2+1d linear SSPT phases.
In particular, setting $I = J$, we find that twice the diagonal response is equivalent to a weak SSPT response.
Thus, the diagonal coefficients satisfy
\begin{align}
    k_{II} \sim k_{II} + 2
\end{align}
under the equivalence relation defining strong SSPT phases.

To establish the nontriviality of these responses as strong SSPT phases, we compactify the system along the $x_2$ direction with circumference $L_2$.
The planar symmetry generators $U_2^{(I)}(x_2)$ then become 0-form symmetries of the dimensionally reduced 2+1d system.
In Appendix~\ref{app:typeII_compactification}, we derive the resulting 2+1d SPT response and evaluate the strong SSPT invariants introduced in Ref.~\cite{Devakul:2019duj}.
We show that these invariants distinguish all independent values of $k_{IJ}$ modulo $\gcd(N_I,N_J)$ and $k_{II}$ modulo $\gcd(N_I,2)$.
Consequently, our response theories realize all strong 2-foliated SSPT classes with Abelian fracton duals
\begin{align}
    \prod_{I<J} \mathbb{Z}_{\gcd(N_I,N_J)} \times \prod_I \mathbb{Z}_{\gcd(N_I,2)},
\end{align}
obtained in Ref.~\cite{Devakul:2019duj}.

\paragraph{Entangler}
Using the action~\eqref{eq:typeII_response_action}, we obtain a corresponding entangler
\begin{align}
    \begin{split}
    V = \sum_{s^{1},\cdots,s^{M}} &\exp\left(2\pi i \sum_{I\leq J}\frac{k_{IJ}}{N_{I}}\sum_{n}s^{I}(n) \cdot \left(\frac{\Delta_{3}[\Delta_{1}\Delta_{2}s^{J}]_{N_{J}}(n) - \Delta_{1}\Delta_{2}[\Delta_{3}s^{J}]_{N_{J}}(n)}{N_{J}}\right)\right)\\
    &\times \ket{\{s^{1}\},\cdots,\{s^{M}\}}\bra{\{s^{1}\},\cdots,\{s^{M}\}}.
    \end{split}
    \label{eq:typeII_entangler}
\end{align}
Here, $[X]_N \in \{0,1,\ldots,N-1\}$ denotes the representative of $X$ modulo $N$.
As shown in Appendix~\ref{app:correspondence_of_entanglers},
when $M = 2$ and $N_1 = N_2 = N$,
the entangler obtained from the response action~\eqref{eq:typeII_response_action}
agrees with that constructed in Ref.~\cite{Devakul:2019duj}
upon an appropriate choice of integer lift in the cochain representative of the Bockstein homomorphism.
While the explicit 2-foliated lattice construction in that reference focuses on this case, our construction applies to arbitrary finite Abelian onsite symmetry groups.

\subsection{2-foliated type-III SSPT in 3+1d}
\label{sec:new_SSPT}
In this subsection, we introduce a new 3+1d SSPT phase protected by two 2-foliated planar $\mathbb{Z}_{N}$ subsystem symmetries and a global $\mathbb{Z}_{N}$ 0-form symmetry.
This SSPT phase is a subsystem analogue of the 2+1d type-III SPT phases protected by three $\mathbb{Z}_{N}$ 0-form symmetries,
with response actions of the form
\begin{align}
    \frac{2\pi i p}{N} \int A\cup B\cup C,
    \label{eq:usual_TypeIII_action}
\end{align}
where $A$, $B$, and $C$ are $\mathbb{Z}_{N}$ gauge fields.
To construct its subsystem counterpart, we introduce a trilinear lattice response action involving three background gauge fields
that goes beyond the intersection pairing considered in Sec.~\ref{sec:intersection_pairing}.

\paragraph{Symmetry action}
We begin with a spin system with three species of qudits on a three-dimensional cubic lattice.
At each site $(x_1,x_2,x_3)=(x,y,z)$, we assign three qudits labeled by $I = a,b,c$.
The generalized Pauli operators acting on the qudit of species $I$ at site $(x,y,z)$ are denoted by $\mathcal{X}_{I}(x,y,z)$ and $\mathcal{Z}_{I}(x,y,z)$.

For each $I=b,c$, we impose a 2-foliated planar $\mathbb{Z}_{N}$ subsystem symmetry generated by
\begin{align}
    \prod_{y,z} \mathcal{X}_{I}(x_0,y,z), \qquad \prod_{x,z} \mathcal{X}_{I}(x,y_0,z).
\end{align}
In addition, we impose a global $\mathbb{Z}_{N}$ 0-form symmetry generated by
\begin{align}
    \prod_{x,y,z} \mathcal{X}_{a}(x,y,z).
\end{align}

\paragraph{Gauge fields}
We denote the background gauge fields for the symmetries acting on qudits of species $a$, $b$, and $c$ by $a$, $b$, and $c$, respectively.

The gauge transformations are given by
\begin{align}
    \begin{split}
    &a_{\mu} \to a_{\mu} + \Delta_{\mu}\alpha,\\
    &b_{0} \to b_{0} + \Delta_{0}\beta, \qquad
    b_{3} \to b_{3} + \Delta_{3}\beta, \qquad
    b_{12} \to b_{12} + \Delta_{1}\Delta_{2}\beta, \\
    &c_{0} \to c_{0} + \Delta_{0}\gamma, \qquad
    c_{3} \to c_{3} + \Delta_{3}\gamma, \qquad
    c_{12} \to c_{12} + \Delta_{1}\Delta_{2}\gamma,
    \end{split}
\end{align}
where $\alpha,\beta,\gamma$ are gauge transformation parameters.
These gauge fields are subject to the following flatness conditions:
\begin{align}
    \begin{split}
    &\Delta_{\mu} a_{\nu} - \Delta_{\nu}a_{\mu} = 0,\\
    &\Delta_{0}b_{3} - \Delta_{3}b_{0} = 0,\qquad
    \Delta_{0}b_{12} - \Delta_{1}\Delta_{2}b_{0} = 0,\qquad
    \Delta_{3}b_{12} - \Delta_{1}\Delta_{2}b_{3} = 0,\\
    &\Delta_{0}c_{3} - \Delta_{3}c_{0} = 0,\qquad
    \Delta_{0}c_{12} - \Delta_{1}\Delta_{2}c_{0} = 0,\qquad
    \Delta_{3}c_{12} - \Delta_{1}\Delta_{2}c_{3} = 0.
    \end{split}
\end{align}
\paragraph{Lattice response action}
The SSPT action is given by
\begin{align}
    \begin{split}
        S_{p} = \frac{2\pi i p}{N}\sum_{n} &\left[ a_{0}(n)\left\{b_{12}(n+\hat{0})c_{3}(n+\hat{0}+\hat{1}+\hat{2}) + b_{3}(n+\hat{0})c_{12}(n+\hat{0}+\hat{3})\right\}\right.\\
        &-a_{1}(n)\left\{\Delta_{2}b_{0}(n+\hat{1})c_{3}(n+\hat{0}+\hat{1}+\hat{2})-\Delta_{2}b_{3}(n+\hat{1})c_{0}(n+\hat{1}+\hat{2}+\hat{3})\right\}\\
        &+a_{2}(n)\left\{b_{0}(n+\hat{2})\Delta_{1}c_{3}(n+\hat{0}+\hat{2})-b_{3}(n+\hat{2})\Delta_{1}c_{0}(n+\hat{2}+\hat{3})\right\}\\
        &\left. -a_{3}(n)\left\{b_{0}(n+\hat{3})c_{12}(n+\hat{0}+\hat{3})+b_{12}(n+\hat{3})c_{0}(n+\hat{1}+\hat{2}+\hat{3})\right\} \right].
    \end{split}
    \label{eq:3+1d_typeIII_SSPT_action}
\end{align}
The exponentiated action is gauge invariant on a closed spacetime lattice,
provided that the flatness conditions are satisfied.
This action provides a subsystem analogue of the conventional type-III response action in Eq.~\eqref{eq:usual_TypeIII_action}.

\paragraph{Lattice model and boundary anomaly}
The entangler for this SSPT phase is given by
\begin{align}
    V = \sum_{s_a,s_b,s_c} e^{i\phi[(s_a,s_b,s_c),0]} \ket{\{s_a\},\{s_b\},\{s_c\}}\bra{\{s_a\},\{s_b\},\{s_c\}},
\end{align}
where
\begin{align}
    \phi[(s_a,s_b,s_c),0] = +\frac{2\pi p}{N} \sum_{x,y,z} s_{a}(x,y,z) \left(\Delta_{1}\Delta_{2} s_{b}(x,y,z) \Delta_{3} s_{c}(x+1,y+1,z)
    + \Delta_{3}s_{b}(x,y,z) \Delta_{1}\Delta_{2} s_{c}(x,y,z+1) \right).
\end{align}
This entangler can be expressed as a product of generalized controlled-controlled-$Z$ (CCZ) gates acting on triples of $\mathbb{Z}_{N}$ qudits.
The corresponding SSPT Hamiltonian is given by
\begin{align}
    H = - V \left\{ \sum_{x,y,z} \left(\mathcal{X}_{a}(x,y,z) + \mathcal{X}_{b}(x,y,z) + \mathcal{X}_{c}(x,y,z)\right) + \mathrm{h.c.} \right\} V^{\dagger}.
\end{align}

To demonstrate the nontrivial SSPT order, we analyze the boundary symmetry action.
We place the SSPT phase in the half-space $z\leq 0$, with a boundary at $z=0$.
Qudits of species $b$ and $c$ are assigned to sites with $z\leq 0$, while those of species $a$ are assigned only to sites with $z<0$.
We impose periodic boundary conditions in the $x$ and $y$ directions.
To construct the bulk Hamiltonian in this geometry, we define a truncated entangler
\begin{align}
    V_{\mathrm{bdry}} = \sum_{s_a,s_b,s_c} e^{i\phi_{\mathrm{bdry}}[s_a,s_b,s_c]} \ket{\{s_a\},\{s_b\},\{s_c\}}\bra{\{s_a\},\{s_b\},\{s_c\}},
\end{align}
where
\begin{align}
    \phi_{\mathrm{bdry}}[s_a,s_b,s_c]
    = +\frac{2\pi p}{N} \sum_{z < 0, x,y} s_{a}(x,y,z) \left(\Delta_{1}\Delta_{2} s_{b}(x,y,z) \Delta_{3} s_{c}(x+1,y+1,z)
    + \Delta_{3}s_{b}(x,y,z) \Delta_{1}\Delta_{2} s_{c}(x,y,z+1) \right).
\end{align}
The bulk Hamiltonian is then given by
\begin{align}
    H_{\mathrm{bulk}} = - g V_{\mathrm{bdry}} \left\{ \sum_{z<0,x,y} \left(\mathcal{X}_{a}(x,y,z) + \mathcal{X}_{b}(x,y,z) + \mathcal{X}_{c}(x,y,z)\right) + \mathrm{h.c.} \right\} V^{\dagger}_{\mathrm{bdry}},
\end{align}
where $g>0$ is taken to be much larger than the energy scale of the boundary physics,
which is specified by a boundary Hamiltonian.
We assume the boundary Hamiltonian preserves the symmetries generated by the operators
\begin{align}
    \prod_{z\leq 0,y} \mathcal{X}_{I}(x_0,y,z),\quad \prod_{z\leq 0,x} \mathcal{X}_{I}(x,y_0,z),\quad \prod_{z < 0,x,y} \mathcal{X}_{a}(x,y,z).
\end{align}
for $I = b,c$.
The remaining low-energy degrees of freedom are effective qudits living on the boundary.
The effective qudit operators are given by
\begin{align}
    \tilde{\mathcal{X}}_{I}(x,y) &= V_{\mathrm{bdry}} \mathcal{X}_{I}(x,y,0) V_{\mathrm{bdry}}^{\dagger},\\
    \tilde{\mathcal{Z}}_{I}(x,y) &= \mathcal{Z}_{I}(x,y,0),
\end{align}
for $I = b,c$.
On the boundary, the bulk symmetry operators become
\begin{align}
    \prod_{z\leq 0,y} \mathcal{X}_{I}(x_0,y,z) &\leadsto \prod_{y} \tilde{\mathcal{X}}_{I}(x_0,y) \quad \text{for } I = b,c, \\
    \prod_{z\leq 0,x} \mathcal{X}_{I}(x,y_0,z) &\leadsto \prod_{x} \tilde{\mathcal{X}}_{I}(x,y_0) \quad \text{for } I = b,c, \\
    \prod_{z < 0,x,y} \mathcal{X}_{a}(x,y,z) &\leadsto \sum_{\tilde{s}_b,\tilde{s}_c}e^{-\frac{2\pi ip}{N}\sum_{x,y}\tilde{s}_{b}(x,y)\Delta_{1}\Delta_{2}\tilde{s}_{c}(x,y)}\ket{\{\tilde{s}_b\},\{\tilde{s}_c\}}\bra{\{\tilde{s}_b\},\{\tilde{s}_c\}},
\end{align}
where $\ket{\{\tilde{s}_b\},\{\tilde{s}_c\}}$ denotes a basis state of the effective boundary qudits.

Thus, the 2+1d boundary supports two linear $\mathbb{Z}_{N}$ subsystem symmetries and a global $\mathbb{Z}_{N}$ 0-form symmetry generated by an entangler for the linear SSPT phase studied in Sec.~\ref{sec:2+1d_linear_SSPT}.
This can also be understood from the final term in Eq.~\eqref{eq:3+1d_typeIII_SSPT_action}, which implies that a codimension-one defect of the 0-form symmetry perpendicular to the $z$-direction is decorated with a linear SSPT.

Suppose that the boundary Hamiltonian admits a unique, symmetric, gapped, short-range-entangled ground state.
Such a state must belong to some SSPT phase protected by the two linear subsystem symmetries.
However, the global $\mathbb{Z}_{N}$ 0-form symmetry acts by stacking a linear SSPT phase labeled by $-p$ modulo $N$.
For $p\neq 0$ modulo $N$, this changes the SSPT class, contradicting the invariance of the ground state under the 0-form symmetry.
This is a manifestation of the 't Hooft anomaly of the boundary theory.

Alternatively, the anomaly can be diagnosed through the projective algebra of symmetry operators in the presence of symmetry defects,
as discussed in Refs.~\cite{Stephen:2020zxm,Li:2022nwa,Furukawa:2025flp,ParayilMana:2025nxw}.\footnote{See Ref.~\cite{Furukawa:2025flp} for a discussion of more general type-III anomalies.}

\subsection{3-foliated type-III SSPT in 3+1d}

In this subsection, we construct a lattice response theory for the 3+1d SSPT phase introduced in Ref.~\cite{Stephen:2020zxm}, where an explicit model was constructed for $N = 2$.
Ref.~\cite{Stephen:2020zxm} also analyzed the anomalous action of symmetries on the boundary.
In particular, the boundary symmetry operators take the same form as those discussed in the previous subsection, despite the different foliation structures of the bulk subsystem symmetries.\footnote{Another example in which SSPT phases protected by symmetries with different foliation structures exhibit the same boundary anomaly was discussed in Ref.~\cite{Burnell:2021reh}.}
Unlike the previous example, the response action constructed here is linear in the ordinary gauge field $A$ and quadratic in a single subsystem gauge field $B$.
We extend the construction to general $\mathbb{Z}_N$ symmetries.

\paragraph{Symmetry action}
We consider a three-dimensional cubic lattice with two species of $\mathbb{Z}_N$ qudits, located on sites and links, respectively.
The generalized Pauli operators acting on the qudit at a site $n$ are denoted by $\mathcal{X}(n)$ and $\mathcal{Z}(n)$.
Those acting on the qudit on the $i$-link originating at $n$ are denoted by $\mathcal{X}_i(n)$ and $\mathcal{Z}_i(n)$.

We impose a global $\mathbb{Z}_N^A$ 0-form symmetry generated by
\begin{align}
    U_A = \prod_n \mathcal{X}(n),
\end{align}
and a 3-foliated $\mathbb{Z}_N^B$ planar subsystem symmetry generated by
\begin{align}
    U_{B,i}^{\perp}(x_i) &= \prod_{x_j,x_k}\mathcal{X}_i(x_1,x_2,x_3),\\
    U_{B,i}^{\parallel}(x_i) &= \prod_{x_j,x_k}\mathcal{X}_j(x_1,x_2,x_3)\mathcal{X}_k^\dagger(x_1,x_2,x_3),
\end{align}
for $(i,j,k) = (1,2,3), (2,3,1), (3,1,2)$, with $x_i$ fixed.

The algebra of local $\mathbb{Z}_N^A$-symmetric operators is generated by $\mathcal{X}(n)$ and
\begin{align}
    \mathcal{Z}_A^i(n) = \mathcal{Z}(n+\hat{i})\mathcal{Z}^\dagger(n).
\end{align}
Similarly, the algebra of local $\mathbb{Z}_N^B$-symmetric operators is generated by $\mathcal{X}_i(n)$ and
\begin{align}
    \mathcal{Z}_B^i(n) &= \mathcal{Z}_j(n+\hat{i})\mathcal{Z}_j^\dagger(n)\mathcal{Z}_k(n+\hat{i})\mathcal{Z}_k^\dagger(n),\\
    \mathcal{Z}_B^{jk}(n) &= \mathcal{Z}_i(n+\hat{j}+\hat{k})\mathcal{Z}_i^\dagger(n+\hat{j})\mathcal{Z}_i^\dagger(n+\hat{k})\mathcal{Z}_i(n),
\end{align}
for cyclic permutations $(i,j,k)$ of $(1,2,3)$.

\paragraph{Gauge fields}
The background gauge field for the $\mathbb{Z}_N^A$ 0-form symmetry is denoted by $A = (A_0,A_1,A_2,A_3)$.
Its gauge transformations and flatness conditions are
\begin{align}
    A_\mu \to A_\mu + \Delta_\mu\alpha,\\
    \Delta_\mu A_\nu - \Delta_\nu A_\mu = 0,
\end{align}
for $\mu,\nu = 0,1,2,3$.

The background gauge field for the $\mathbb{Z}_N^B$ subsystem symmetry is denoted by $B = (B_{0i},B_i,B_{jk})$.
Its gauge transformations are
\begin{align}
    B_{0i} &\to B_{0i} + \Delta_{0} \beta_{i}, \\
    B_{i} &\to B_{i} + \Delta_{i}\beta_{j} + \Delta_{i}\beta_{k},\\
    B_{jk} &\to B_{jk} + \Delta_{j}\Delta_{k}\beta_{i},
\end{align}
and its flatness conditions are
\begin{align}
    \Delta_{0} B_{i} - \Delta_{i} B_{0j} - \Delta_{i} B_{0k} &= 0,\\
    \Delta_{0}B_{jk} - \Delta_{j}\Delta_{k}B_{0i} &= 0,\\
    \Delta_{j}\Delta_{k} B_{i} - \Delta_{j} B_{ki} - \Delta_{k} B_{ij} &= 0,
\end{align}
for cyclic permutations $(i,j,k)$ of $(1,2,3)$.

\paragraph{SSPT response action}
A flat $\mathbb{Z}_{N}^{B}$ gauge field satisfies
\begin{align}
    \begin{split}
    &\Delta_{0} \sum_{(ijk)} \left(B_{jk}(n)(B_{ki}+B_{i})(n)\right)\\
    =& \sum_{(ijk)} \Delta_{i}\left\{(B_{0j}+B_{0k})(n)B_{jk}(n-\hat{i})\right.\\
    &\left. + \Delta_{j}B_{0k}(n)(B_{jk}+B_{ki}+B_{k})(n-\hat{i}) - B_{0j}(n)\Delta_{k}(B_{ij}+B_{jk}+B_{j})(n-\hat{k}-\hat{i})\right.\\
    &\left. +B_{0k}(n)\Delta_{j}\Delta_{k}B_{0i}(n-\hat{i}) + \Delta_{j}B_{0k}(n)\Delta_{k}B_{0i}(n-\hat{i}+\hat{j}) - B_{0j}(n)\Delta_{j}\Delta_{k}B_{0k}(n-\hat{k}) \right\},
    \end{split}
\end{align}
where $\sum_{(ijk)}$ denotes the sum over cyclic permutations of $(1,2,3)$.
Combining this identity with the ordinary gauge field $A$, we construct the following gauge-invariant response action:
\begin{align}
    \begin{split}
    S_{p} = &\frac{2\pi ip}{N} \sum_{n} \left[A_{0}(n)\left(\sum_{(ijk)} B_{jk}(n+\hat{0})(B_{ki}+B_{i})(n+\hat{0})\right)\right. \\
    &- \sum_{(ijk)} A_{i}(n)\left\{B_{0j}(n+\hat{i})B_{jk}(n) + B_{0k}(n+\hat{i})B_{jk}(n)\right.\\
    &+ \Delta_{j}B_{0k}(n+\hat{i})(B_{jk}+B_{ki}+B_{k})(n) - B_{0j}(n+\hat{i})\Delta_{k}(B_{ij}+B_{jk}+B_{j})(n-\hat{k}) \\
    &\left.\left. +B_{0k}(n+\hat{i})\Delta_{j}\Delta_{k}B_{0i}(n) + \Delta_{j}B_{0k}(n+\hat{i})\Delta_{k}B_{0i}(n+\hat{j}) - B_{0j}(n+\hat{i})\Delta_{j}\Delta_{k}B_{0k}(n-\hat{k}+\hat{i})\right\}\right],
    \end{split}
\end{align}
where $p\in\mathbb{Z}_N$.

For the following flat gauge field configuration on a periodic spacetime lattice,
\begin{align}
    A_0(x_0,x_1,x_2,x_3) &= \delta(x_0),\qquad A_i(x_0,x_1,x_2,x_3) = 0,\\
    B_1(x_0,x_1,x_2,x_3) &= \delta(x_1)\delta(x_3),\qquad B_{23}(x_0,x_1,x_2,x_3) = \delta(x_2),\\
    B_2 &= B_3 = B_{12} = B_{31} = B_{0i} = 0,
\end{align}
where $i = 1,2,3$, we obtain
\begin{align}
    S_p[A,B] = \frac{2\pi ip}{N}\mod 2\pi i.
\end{align}
Thus, the response action is nontrivial for $p\neq 0$ and distinguishes all $p\in\mathbb{Z}_N$.

\paragraph{Entangler}
Using the method in Sec.~\ref{sec:entangler}, we obtain the entangler
\begin{align}
    V_{p} =
    \sum_{s,s_{1},s_{2},s_{3}}
    e^{\frac{2\pi ip}{N}\sum_{n}s(n)\left(\sum_{(ijk)} \Delta_{j}\Delta_{k}s_{i}(n)(\Delta_{k}\Delta_{i}s_{j} + \Delta_{i}s_{j} + \Delta_{i}s_{k})(n)\right)}
    \ket{\{s\},\{s_{1}\},\{s_{2}\},\{s_{3}\}}\bra{\{s\},\{s_{1}\},\{s_{2}\},\{s_{3}\}}.
\end{align}
For $N = 2$ and $p = 1$, this reproduces the CCZ entangler of Ref.~\cite{Stephen:2020zxm} upon identifying the site qudits with the qubits at the centers of cubes.
Thus, our construction extends the lattice realization of Ref.~\cite{Stephen:2020zxm} to general $\mathbb{Z}_N$ symmetries.

\section{Conclusion}

In this work, we constructed lattice response theories describing various SSPT phases,
based on the systematic framework developed in Sec.~\ref{sec:lattice_gauge_fields} for constructing lattice gauge fields from symmetry actions and entanglers from given gauge-invariant response actions.
We showed that intersection pairings yield a broad class of SSPT phases realized by cluster models in Sec.~\ref{sec:intersection_pairing}.
We further studied SSPT phases beyond intersection pairings in Sec.~\ref{sec:beyond_intersection}.
In particular, as discussed in Sec.~\ref{sec:new_SSPT}, we constructed new SSPT phases described by subsystem analogues of type-III group cocycles.

We conclude by discussing several future directions.
An important open problem is to develop a systematic construction of gauge-invariant response actions beyond intersection pairings.
We expect this problem to be closely related to the construction of logical gates in quantum error-correcting codes.
In particular, for Calderbank-Shor-Steane (CSS) codes described by (co)chain complexes,
the code space has a basis labeled by (co)homology classes, and certain logical gates can be constructed using cohomological operations.
See Refs.~\cite{Breuckmann:2024axh,Hsin:2024nwc,Li:2026cwn,Sahay:2026yps,Breuckmann:2026pkw} for related developments.

Another interesting direction is to extend our framework to fermionic systems and time-reversal symmetry.
See Refs.~\cite{You:2018oai,You:2018zhj,Burnell:2021reh,Cao:2022lig,Zhang:2022wik} for related discussions.

A systematic classification of strong SSPT phases is also an important open problem.
Although classifications have been established for particular classes of subsystem symmetry actions~\cite{Devakul:2018fhz,Devakul:2019duj},
it would be interesting to develop a general algorithm that determines the classification of strong SSPT phases from a given microscopic symmetry action.

\section*{Acknowledgments}
Y.F. was supported by JST SPRING, Grant Number JPMJSP2108.
T.O. was supported in part by JST PRESTO, Grant Number JPMJPR23F3.

ChatGPT (GPT-6) was used for language editing, literature searches, discussions of theoretical constructions, and preliminary consistency checks.
Claude (Fable 5.2 and Opus 5.5) was also used to assist in checking mathematical derivations.
The lattice response theories presented in this work were derived and verified by the authors.
The authors take full responsibility for the theoretical results and conclusions presented in this work.

\appendix

\section{Relative translations of gauge fields and weak SSPT responses}
\label{app:weak_SSPT}

As discussed in Sec.~\ref{sec:2+1d_linear_SSPT},
we show that finite relative translations of the background gauge fields do not change the strong SSPT class.
To this end, we consider the response action
\begin{align}
    -k_{AB} S_{V}[A,B] = \frac{2\pi i k_{AB}}{\gcd(N_{A},N_{B})} \sum_{n} (A_{0}(n)B_{12}(n+\hat{0}+V)+A_{12}(n)B_{0}(n+\hat{1}+\hat{2}+V)),
\end{align}
where $V$ is a finite spacetime shift vector for the $\mathbb{Z}_{N_B}$ gauge field $B$.
Throughout this appendix, we assume periodic boundary conditions in spacetime and restrict to flat background gauge fields.
We show that the difference
\begin{align}
    \Delta S_{\mu} = S_{\hat{\mu}} - S_0
\end{align}
vanishes for temporal translations and can be decomposed into lower-dimensional SPT response actions for spatial translations.

\paragraph{$\mu = 0$}
For flat $A$ and $B$, we have
\begin{align}
    \begin{split}
    \quad i \Delta S_{0} &= \frac{2\pi}{\gcd(N_{A},N_{B})} \sum_{n} (A_{0}(n)\Delta_{0}B_{12}(n+\hat{0})+A_{12}(n)\Delta_{0}B_{0}(n+\hat{1}+\hat{2})) \\
    &= \frac{2\pi}{\gcd(N_{A},N_{B})} \sum_{n} (A_{0}(n)\Delta_{0}B_{12}(n+\hat{0}) - \Delta_{0}A_{12}(n) B_{0}(n+\hat{0}+\hat{1}+\hat{2})) \\
    &= \frac{2\pi}{\gcd(N_{A},N_{B})} \sum_{n} (A_{0}(n)\Delta_{0}B_{12}(n+\hat{0}) - \Delta_{1}\Delta_{2} A_{0}(n) B_{0}(n+\hat{0}+\hat{1}+\hat{2})) \\
    &= \frac{2\pi}{\gcd(N_{A},N_{B})} \sum_{n} A_{0}(n)(\Delta_{0}B_{12}(n+\hat{0}) - \Delta_{1}\Delta_{2} B_{0}(n+\hat{0})) \\
    &= 0.
    \end{split}
\end{align}

\paragraph{$\mu = 1$}

We have
\begin{align}
    \begin{split}
    i \Delta S_{1}
    &= \frac{2\pi}{\gcd(N_{A},N_{B})} \sum_{n}\left( A_{0}(n)\Delta_{1}B_{12}(n+\hat{0}) + A_{12}(n) \Delta_{1} B_{0}(n+\hat{1}+\hat{2}) \right) \\
    &= -\frac{2\pi}{\gcd(N_{A},N_{B})} \sum_{n}\left( \Delta_{1}A_{0}(n)B_{12}(n+\hat{0}+\hat{1}) - A_{12}(n) \Delta_{1} B_{0}(n+\hat{1}+\hat{2}) \right) \\
    &= -\sum_{x_1} I_{x_{1}},
    \end{split}
\end{align}
where
\begin{align}
    I_{x_{1}}
    = \frac{2\pi}{\gcd(N_{A},N_{B})} \sum_{x_{0},x_{2}}\left( \Delta_{1}A_{0}(x_0,x_1,x_2)B_{12}(x_0+1,x_1+1,x_2) - A_{12}(x_0,x_1,x_2) \Delta_{1} B_{0}(x_0,x_1+1,x_2+1) \right).
\end{align}
Each $I_{x_1}$ is gauge invariant modulo $2\pi$ and defines the phase of a 1+1d SPT response protected by $\mathbb{Z}_{N_A}\times\mathbb{Z}_{N_B}$ 0-form symmetries.
The corresponding gauge fields on the $(x_0,x_2)$ plane labeled by $x_1$ are
$(\tilde{A}_{0}^{(x_1)}(x_0,x_2),\tilde{A}_{2}^{(x_1)}(x_0,x_2)) = (\Delta_{1}A_{0}(x_0,x_1,x_2), A_{12}(x_0,x_1,x_2))$
and $(\tilde{B}_{0}^{(x_1)}(x_0,x_2),\tilde{B}_{2}^{(x_1)}(x_0,x_2)) = (\Delta_{1}B_{0}(x_0,x_1+1,x_2), B_{12}(x_0,x_1+1,x_2))$.
These fields satisfy the usual 1+1d gauge transformations and flatness conditions.
Consequently, we can write
\begin{align}
    I_{x_1} = \frac{2\pi}{\gcd(N_A,N_B)} \int_{M_{x_1}} \tilde{A}^{(x_1)}\cup\tilde{B}^{(x_1)},
\end{align}
where $M_{x_1}$ denotes the $(x_0,x_2)$ plane labeled by $x_1$,
and the cup product is understood modulo $\gcd(N_A,N_B)$.
Thus, the difference between the original and translated response actions is a sum of 1+1d SPT responses,
corresponding to the stacking of weak SSPT phases.

\paragraph{General translations}
The same argument applies to a translation along the $x_2$ direction.
In this case, $\Delta S_2$ can be expressed as a sum of 1+1d SPT response actions on the $(x_0,x_1)$ planes at fixed $x_2$.
An arbitrary finite spacetime translation can be decomposed into elementary translations along the coordinate directions $\mu=0,1,2$.
Since temporal translations leave the response unchanged and spatial translations change it only by stacking with lower-dimensional SPT phases,
the strong SSPT class is invariant under such translations.

\section{Compactification of the Bockstein response action}
\label{app:typeII_compactification}

In this appendix, following the dimensional-reduction approach of Ref.~\cite{Devakul:2019duj}, we compactify the SSPT phases~\eqref{eq:typeII_response_action}
protected by 2-foliated planar subsystem symmetries along the $x_2$ direction with circumference $L_2$.
We derive the response action of the resulting 2+1d system,
which has the 0-form symmetry group $(\mathbb{Z}_{N_1}\times\cdots\times\mathbb{Z}_{N_M})^{L_2}$ generated by
\begin{align}
    U_{2}^{(I)}(x_2) = \prod_{x_1,x_3} \mathcal{X}^{(I)}_{(x_1,x_2,x_3)}.
\end{align}

We denote the lattice gauge field associated with a $\mathbb{Z}_{N_{I}}$ symmetry generated by $U_{2}^{(I)}(x_{2})$
as $\tilde{A}^{(I,x_2)}$.
We assume that the gauge fields satisfy the flatness conditions
\begin{align}
    \Delta_{\mu}\tilde{A}^{(I,x_2)}_{\nu} - \Delta_{\nu}\tilde{A}^{(I,x_2)}_{\mu} = 0 \mod N_{I},
\end{align}
for $\mu,\nu \in \{0,1,3\}$.
The embedding of the 0-form symmetry gauge fields induces
\begin{align}
    A^I_{0}(x_0,x_1,x_2,x_3) &= \tilde{A}^{(I,x_2)}_{0}(x_0,x_1,x_3),\\
    A^I_{12}(x_0,x_1,x_2,x_3) &= \Delta_{2} \tilde{A}^{(I,x_2)}_{1}(x_0,x_1,x_3), \\
    A^I_{3}(x_0,x_1,x_2,x_3) &= \tilde{A}^{(I,x_2)}_{3}(x_0,x_1,x_3).
\end{align}

Substituting these into Eq.~\eqref{eq:typeII_response_action},
we obtain
\begin{align}
    \begin{split}
    -S = \sum_{n}\sum_{I\leq J} \frac{2\pi i k_{IJ}}{N_{I}N_{J}} \left(
        [A^I_{0}(n)]_{N_{I}} (\Delta_{3}[A^J_{12}]_{N_{J}}-\Delta_{1}\Delta_{2}[A_{3}^J]_{N_{J}})(n+\hat{0}) \right.\\
        -[A^I_{3}(n)]_{N_{I}} (\Delta_{0}[A^J_{12}]_{N_{J}}-\Delta_{1}\Delta_{2}[A_{0}^J]_{N_{J}})(n+\hat{3})\\
        \left.-[A^I_{12}(n)]_{N_{I}} (\Delta_{0}[A^J_{3}]_{N_{J}}-\Delta_{3}[A_{0}^J]_{N_{J}})(n+\hat{1}+\hat{2})
    \right)\\
    = \sum_{x_{0},x_{1},x_{2},x_{3}}\sum_{I\leq J} \frac{2\pi i k_{IJ}}{N_{I}N_{J}} \left\{
        [\tilde{A}^{(I,x_2)}_{0}(x_0,x_1,x_3)]_{N_{I}} (\Delta_{3}[\Delta_{2} \tilde{A}^{(J,x_2)}_{1}]_{N_{J}}-\Delta_{1}\Delta_{2}[\tilde{A}^{(J,x_2)}_{3}]_{N_{J}})(x_{0}+1,x_{1},x_{3}) \right.\\
        -[\tilde{A}^{(I,x_2)}_{3}(x_0,x_1,x_3)]_{N_{I}} (\Delta_{0}[\Delta_{2} \tilde{A}^{(J,x_2)}_{1}]_{N_{J}}-\Delta_{1}\Delta_{2}[\tilde{A}^{(J,x_2)}_{0}]_{N_{J}})(x_{0},x_{1},x_{3}+1)\\
        \left.-[\Delta_{2} \tilde{A}^{(I,x_2)}_{1}(x_0,x_1,x_3)]_{N_{I}} (\Delta_{0}[\tilde{A}^{(J,x_2+1)}_{3}]_{N_{J}}-\Delta_{3}[\tilde{A}^{(J,x_2+1)}_{0}]_{N_{J}})(x_{0},x_{1}+1,x_{3})
    \right\} \\
    = \sum_{x_{0},x_{1},x_{2},x_{3}}\sum_{I\leq J} \frac{2\pi i k_{IJ}}{N_{I}N_{J}} \left\{
        [\tilde{A}^{(I,x_2)}_{0}(x_0,x_1,x_3)]_{N_{I}} (\Delta_{3}\Delta_{2} [\tilde{A}^{(J,x_2)}_{1}]_{N_{J}}-\Delta_{1}\Delta_{2}[\tilde{A}^{(J,x_2)}_{3}]_{N_{J}})(x_{0}+1,x_{1},x_{3}) \right.\\
        -[\tilde{A}^{(I,x_2)}_{3}(x_0,x_1,x_3)]_{N_{I}} (\Delta_{0}\Delta_{2} [\tilde{A}^{(J,x_2)}_{1}]_{N_{J}}-\Delta_{1}\Delta_{2}[\tilde{A}^{(J,x_2)}_{0}]_{N_{J}})(x_{0},x_{1},x_{3}+1)\\
        + [\tilde{A}^{(I,x_2)}_{1}(x_0,x_1,x_3)]_{N_{I}} \Delta_{2}(\Delta_{0}[\tilde{A}^{(J,x_2)}_{3}]_{N_{J}}-\Delta_{3}[\tilde{A}^{(J,x_2)}_{0}]_{N_{J}})(x_{0},x_{1}+1,x_{3}) \\
        -\Delta_{2} \left([\tilde{A}^{(I,x_2)}_{1}(x_0,x_1,x_3)]_{N_{I}} (\Delta_{0}[\tilde{A}^{(J,x_2)}_{3}]_{N_{J}}-\Delta_{3}[\tilde{A}^{(J,x_2)}_{0}]_{N_{J}})(x_{0},x_{1}+1,x_{3})\right) \\
        + \Delta_{3} \left([\tilde{A}^{(I,x_2)}_{0}(x_0,x_1,x_3)]_{N_{I}} ([\Delta_{2} \tilde{A}^{(J,x_2)}_{1}]_{N_{J}} - \Delta_{2} [\tilde{A}^{(J,x_2)}_{1}]_{N_{J}})(x_{0}+1,x_{1},x_{3})\right)\\
        - \Delta_{0} \left([\tilde{A}^{(I,x_2)}_{3}(x_0,x_1,x_3)]_{N_{I}} ([\Delta_{2} \tilde{A}^{(J,x_2)}_{1}]_{N_{J}} - \Delta_{2} [\tilde{A}^{(J,x_2)}_{1}]_{N_{J}})(x_{0},x_{1},x_{3}+1)\right)\\
        + (\Delta_{0} [\tilde{A}^{(I,x_2)}_{3}]_{N_{I}} - \Delta_{3} [\tilde{A}^{(I,x_2)}_{0}]_{N_{I}})(x_0,x_1,x_3) ([\Delta_{2} \tilde{A}^{(J,x_2)}_{1}]_{N_{J}} - \Delta_{2} [\tilde{A}^{(J,x_2)}_{1}]_{N_{J}})(x_{0}+1,x_{1},x_{3}+1)\\
        \left.
        +(\Delta_{2}[ \tilde{A}^{(I,x_2)}_{1}(x_0,x_1,x_3)]_{N_{I}}-[\Delta_{2} \tilde{A}^{(I,x_2)}_{1}(x_0,x_1,x_3)]_{N_{I}})  (\Delta_{0}[\tilde{A}^{(J,x_2+1)}_{3}]_{N_{J}}-\Delta_{3}[\tilde{A}^{(J,x_2+1)}_{0}]_{N_{J}})(x_{0},x_{1}+1,x_{3})
    \right\}.
    \end{split}
\end{align}
On a closed spacetime, the total difference terms do not contribute,
while the last two terms vanish modulo $2\pi i$ by the flatness conditions.
We thus obtain the dimensionally reduced response action
\begin{align}
    -S = \sum_{x_2}\sum_{I\leq J}\frac{2\pi i k_{IJ}}{N_I}\int_{M^3}\tilde{A}^{(I,x_2)}\cup\left(\Bock_{N_J,N_I}(\tilde{A}^{(J,x_2+1)})-\Bock_{N_J,N_I}(\tilde{A}^{(J,x_2)})\right),
    \label{eq:reduced_response_action}
\end{align}
where $M^3$ denotes the dimensionally reduced spacetime and $\Bock_{N_J,N_I}$ is the Bockstein homomorphism
\begin{align}
    \Bock_{N_J,N_I}: H^1(M^3,\mathbb{Z}_{N_J}) \to H^2(M^3,\mathbb{Z}_{N_I})
\end{align}
associated with the short exact sequence
\begin{align}
    0 \longrightarrow \mathbb{Z}_{N_I} \xrightarrow{\times N_J} \mathbb{Z}_{N_IN_J} \xrightarrow{\operatorname{mod} N_J} \mathbb{Z}_{N_J} \longrightarrow 0.
\end{align}

\paragraph{Strong SSPT invariants}
We compute the strong SSPT invariants introduced in Ref.~\cite{Devakul:2019duj}.
The type-I and type-II cohomology invariants, $p_{\mathrm{I}}$ and $p_{\mathrm{II}}$, of the dimensionally reduced 2+1d SPT are encoded in a symmetric matrix $M$ defined by
\begin{align}
    M_{aa} = 2p_{\mathrm{I}}^a,\qquad M_{ab} = p_{\mathrm{II}}^{ab} = M_{ba}\quad (a\neq b),
\end{align}
where $a = (I,z)$ labels the generators of the 0-form symmetry.
The diagonal entries $M_{aa}$ are even integers defined modulo $2N_I$,
while the off-diagonal entries $M_{ab}$ are defined modulo $\gcd(N_I,N_J)$ for $a = (I,z)$ and $b = (J,z')$.
The strong SSPT invariants $F_1^I$ and $F_2^{IJ}$ are defined by
\begin{align}
    F_1^I &= \sum_{z<z_0}\sum_{z'\geq z_0} M_{(I,z),(I,z')} \mod 2,\\
    F_2^{IJ} &= \sum_{z<z_0}\sum_{z'\geq z_0} \left(M_{(I,z),(J,z')} - M_{(J,z),(I,z')}\right) \mod\gcd(N_I,N_J),
\end{align}
where $F_1^I$ is defined for even $N_I$ and $F_2^{IJ}$ for $I<J$.
These invariants are unchanged by stacking lower-dimensional SPT phases.
To define these invariants, we take the limit $L_2\to\infty$ and consider a single cut at $z=z_0$ in the resulting infinite stack of planes.

For the reduced response action~\eqref{eq:reduced_response_action}, we obtain
\begin{align}
    M_{(I,z),(J,z')} =
    \begin{cases}
        k_{II}(\delta_{z+1,z'}+\delta_{z-1,z'}-2\delta_{z,z'}), & I=J,\\
        k_{IJ}(\delta_{z+1,z'}-\delta_{z,z'}), & I<J,\\
        k_{JI}(\delta_{z-1,z'}-\delta_{z,z'}), & I>J.
    \end{cases}
\end{align}
Consequently,
\begin{align}
    F_1^I = k_{II} \mod 2,\qquad F_2^{IJ} = k_{IJ} \mod \gcd(N_I,N_J).
\end{align}
Since the coefficients $k_{II}$ and $k_{IJ}$ can be chosen independently,
the response theory~\eqref{eq:typeII_response_action} realizes all possible values of the strong SSPT invariants classified in Ref.~\cite{Devakul:2019duj}.

\section{Comparison with the Devakul-Shirley-Wang construction}
\label{app:correspondence_of_entanglers}

In this appendix, we show that the response action~\eqref{eq:typeII_response_action} reproduces the entangler constructed in Ref.~\cite{Devakul:2019duj} when $M=2$ and $N_{1}=N_{2}=N$.

In our convention,\footnote{We identify $(x,y,z)_{\mathrm{DSW}} = (x_2,x_3,x_1)$,
$g^\alpha_{\mathbf r} = s^I(n)$ with $\alpha = I$ and
$\mathbf r = (n_2,n_3,n_1)$, and $q^{\alpha\beta} = -k_{IJ}$
with $(\alpha,\beta) = (I,J)$.} the entangler proposed in Appendix G of Ref.~\cite{Devakul:2019duj} is given by
\begin{align}
    V_{\mathrm{DSW}} = \sum_{s^{1},s^{2}} e^{-i\phi_{\mathrm{DSW}}[s^{1},s^{2}]}\ket{\{s^{1}\},\{s^{2}\}}\bra{\{s^{1}\},\{s^{2}\}},
\end{align}
where
\begin{align}
    \begin{split}
    \phi_{\mathrm{DSW}}[s^{1},s^{2}] &= \sum_{n}\sum_{I\leq J}\frac{2\pi k_{IJ}}{N^2}\left\{
        \Delta_{1}s^{I}(n)\left([\Delta_{3}s^J](n+\hat{1}) + [\Delta_{2}s^J](n+\hat{1}+\hat{3}) - [\Delta_{2}s^J](n+\hat{1}) - [\Delta_{3}s^J](n+\hat{1}+\hat{2})\right)\right.\\
        &\qquad \left.- \Delta_{3} s^{I}(n) \left([\Delta_{2}s^J](n+\hat{1}+\hat{3}) - [\Delta_{2}s^J](n+\hat{3}) + [-\Delta_{1}\Delta_{2}s^J](n+\hat{3}) \right)
    \right\}.
    \end{split}
\end{align}
Here, $[X] \in \{0,1,\ldots,N-1\}$ denotes the representative of $X$ modulo $N$.
Then, we have
\begin{align}
    \begin{split}
    \phi_{\mathrm{DSW}}[s^{1},s^{2}] &= \sum_{n}\sum_{I\leq J}\frac{2\pi k_{IJ}}{N^2}\left\{
        \Delta_{1}s^{I}(n)\left(-\Delta_{2}[\Delta_{3}s^J](n+\hat{1}) + \Delta_{3}[\Delta_{2}s^J](n+\hat{1})\right)\right.\\
        &\qquad \left.- \Delta_{3} s^{I}(n) \left(\Delta_{1}[\Delta_{2}s^J](n+\hat{3}) + [-\Delta_{1}\Delta_{2}s^J](n+\hat{3}) \right)
    \right\} \\
    &= \sum_{n}\sum_{I\leq J}\frac{2\pi k_{IJ}}{N^2}\left\{
        \Delta_{1}(s^{I}\left(-\Delta_{2}[\Delta_{3}s^J] + \Delta_{3}[\Delta_{2}s^J])\right)(n) - \Delta_{3} (s^{I} \left(\Delta_{1}[\Delta_{2}s^J] + [-\Delta_{1}\Delta_{2}s^J])\right)(n) \right.\\
        &\qquad \left. + s^{I} (\Delta_{1}\Delta_{2}[\Delta_{3}s^J] + \Delta_{3} [-\Delta_{1}\Delta_{2}s^J])(n)
    \right\}
    \end{split}
\end{align}

The Bockstein homomorphism is independent of the choice of integer lift,
but its cochain representative depends on this choice.
To reproduce the entangler $V_{\mathrm{DSW}}$, we choose the integer lifts
\begin{align}
    \widetilde{A}_0 = -[-A_0],\qquad
    \widetilde{A}_3 = -[-A_3],\qquad
    \widetilde{A}_{12} = [A_{12}].
\end{align}
The corresponding cochain representative of the Bockstein homomorphism is
\begin{align}
    (\Bock' A)_{03}(n) &= \frac{-\Delta_0[-A_3](n) + \Delta_3[-A_0](n)}{N} \mod N,\\
    (\Bock' A)_{012}(n) &= \frac{\Delta_0[A_{12}](n) + \Delta_1\Delta_2[-A_0](n)}{N} \mod N,\\
    (\Bock' A)_{123}(n) &= \frac{\Delta_3[A_{12}](n) + \Delta_1\Delta_2[-A_3](n)}{N} \mod N.
\end{align}

Using this, the response action gives an entangler
\begin{align}
    V = \sum_{s^1,s^2} e^{-i\sum_n\sum_{I\le J}\frac{2\pi k_{IJ}}{N^2}s^I(\Delta_1\Delta_2[\Delta_3s^J] + \Delta_3[-\Delta_1\Delta_2s^J])(n)} \ket{\{s^1\},\{s^2\}}\bra{\{s^1\},\{s^2\}},
\end{align}
which reproduces $V_{\mathrm{DSW}}$ under periodic spatial boundary conditions.

\bibliographystyle{JHEP}
\bibliography{sspt_from_chain_complex.bib}

\end{document}

%% file: sspt.tikzstyles
\tikzstyle{pointoperator}=[fill=black, draw=none, shape=circle, minimum size=.1 cm, inner sep=0 pt]
\tikzstyle{red dot}=[fill=red, draw=red, shape=circle, minimum size=.1 cm, inner sep=0 pt]
\tikzstyle{red diamond}=[fill=red, draw=red, shape=diamond]
\tikzstyle{blue dot}=[fill=blue, draw=blue, shape=circle]

\tikzstyle{dashedline}=[dashed, -, thick]
\tikzstyle{blueline}=[-, draw=blue, thick]
\tikzstyle{arrowline}=[<-, thick]
\tikzstyle{normalline}=[-, thick, draw={rgb,255: red,171; green,171; blue,171}]
\tikzstyle{fillline}=[-, thick, fill=red, fill opacity=.5]
\tikzstyle{bluefillline}=[-, fill opacity=.5, fill=blue, thick]
\tikzstyle{purplefillline}=[-, fill opacity=.5, fill=purple, thick]
\tikzstyle{bluedashedline}=[-, draw=blue, thick, dashed]

%% file: figures/double-lin-sspt.tikz
\begin{tikzpicture}
	\begin{pgfonlayer}{nodelayer}
		\node [style=none] (0) at (0, 0) {$\mathcal{X}_{A}$};
		\node [style=none] (1) at (0, 2) {};
		\node [style=none] (2) at (2, 2) {};
		\node [style=none] (3) at (2, 0) {};
		\node [style=none] (4) at (-2, 0) {};
		\node [style=none] (5) at (-2, -2) {};
		\node [style=none] (6) at (0, -2) {};
		\node [style=none] (7) at (-2, 2) {};
		\node [style=none] (8) at (2, -2) {};
		\node [style=none] (9) at (-1, 1) {$\mathcal{Z}_{B}$};
		\node [style=none] (10) at (1, 1) {$\mathcal{Z}_{B}$};
		\node [style=none] (11) at (1, -1) {$\mathcal{Z}_{B}$};
		\node [style=none] (12) at (-1, -1) {$\mathcal{Z}_{B}$};
	\end{pgfonlayer}
	\begin{pgfonlayer}{edgelayer}
		\draw [style=normalline] (1.center) to (0.center);
		\draw [style=normalline] (0.center) to (3.center);
		\draw [style=normalline] (3.center) to (2.center);
		\draw [style=normalline] (2.center) to (1.center);
		\draw [style=normalline] (0.center) to (6.center);
		\draw [style=normalline] (6.center) to (5.center);
		\draw [style=normalline] (5.center) to (4.center);
		\draw [style=normalline] (4.center) to (0.center);
		\draw [style=normalline] (3.center) to (8.center);
		\draw [style=normalline] (8.center) to (6.center);
		\draw [style=normalline] (7.center) to (1.center);
		\draw [style=normalline] (7.center) to (4.center);
	\end{pgfonlayer}
\end{tikzpicture}

%% file: figures/double-lin-sspt-sononi.tikz
\begin{tikzpicture}
	\begin{pgfonlayer}{nodelayer}
		\node [style=none] (0) at (-1, 1) {$\mathcal{Z}_{A}$};
		\node [style=none] (1) at (-1, -1) {$\mathcal{Z}_{A}$};
		\node [style=none] (2) at (1, -1) {$\mathcal{Z}_{A}$};
		\node [style=none] (3) at (1, 1) {$\mathcal{Z}_{A}$};
		\node [style=none] (4) at (0, 0) {$\mathcal{X}_{B}$};
	\end{pgfonlayer}
	\begin{pgfonlayer}{edgelayer}
		\draw [style=normalline] (1.center) to (0.center);
		\draw [style=normalline] (0.center) to (3.center);
		\draw [style=normalline] (3.center) to (2.center);
		\draw [style=normalline] (2.center) to (1.center);
	\end{pgfonlayer}
\end{tikzpicture}

%% file: figures/single-lin-sspt.tikz
\begin{tikzpicture}
	\begin{pgfonlayer}{nodelayer}
		\node [style=none] (0) at (0, 0) {$\mathcal{X}_{A}$};
		\node [style=none] (1) at (0, 2) {$\mathcal{Z}_{A}$};
		\node [style=none] (2) at (2, 2) {$\mathcal{Z}_{A}$};
		\node [style=none] (3) at (2, 0) {$\mathcal{Z}_{A}$};
		\node [style=none] (4) at (-2, 0) {$\mathcal{Z}_{A}$};
		\node [style=none] (5) at (-2, -2) {$\mathcal{Z}_{A}$};
		\node [style=none] (6) at (0, -2) {$\mathcal{Z}_{A}$};
		\node [style=none] (7) at (-2, 2) {};
		\node [style=none] (8) at (2, -2) {};
	\end{pgfonlayer}
	\begin{pgfonlayer}{edgelayer}
		\draw [style=normalline] (1.center) to (0.center);
		\draw [style=normalline] (0.center) to (3.center);
		\draw [style=normalline] (3.center) to (2.center);
		\draw [style=normalline] (2.center) to (1.center);
		\draw [style=normalline] (0.center) to (6.center);
		\draw [style=normalline] (6.center) to (5.center);
		\draw [style=normalline] (5.center) to (4.center);
		\draw [style=normalline] (4.center) to (0.center);
		\draw [style=normalline] (3.center) to (8.center);
		\draw [style=normalline] (8.center) to (6.center);
		\draw [style=normalline] (7.center) to (1.center);
		\draw [style=normalline] (7.center) to (4.center);
	\end{pgfonlayer}
\end{tikzpicture}